\documentclass{emulateapj}

\slugcomment{ApJ, in press}

\shorttitle{Heavily-Obscured AGN in Star-Forming Galaxies}
\shortauthors{Treister et al.}

\begin{document}

\title{Heavily Obscured AGN in Star-Forming Galaxies at $z$$\simeq$2\footnote{Partly based on observations collected at the European
Southern Observatory, Chile, under program 080.A-0612.}}

\author{E. Treister\altaffilmark{1,2}, Carolin N. Cardamone\altaffilmark{3,4}, Kevin Schawinski\altaffilmark{2,3,4}, C. Megan Urry\altaffilmark{3,4,5}, Eric Gawiser\altaffilmark{6}, Shanil Virani\altaffilmark{3,4}, 
Paulina Lira\altaffilmark{7}, Jeyhan Kartaltepe\altaffilmark{1}, Maaike Damen\altaffilmark{8}, Edward 
N. Taylor\altaffilmark{8}, Emeric  Le Floc'h\altaffilmark{1}, Stephen Justham\altaffilmark{9}, Anton M. Koekemoer\altaffilmark{10}}

\altaffiltext{1}{Institute for Astronomy, 2680 Woodlawn Drive, University of Hawaii, Honolulu, HI 96822; treister@ifa.hawaii.edu}
\altaffiltext{2}{Chandra/Einstein Fellow}
\altaffiltext{3}{Department of Astronomy, Yale University, PO Box 208101, New Haven, CT 06520.}
\altaffiltext{4}{Yale Center for Astronomy and Astrophysics, P.O. Box 208121, New Haven, CT 06520.}
\altaffiltext{5}{Department of Physics, Yale University, P.O. Box 208121, New Haven, CT 06520.}
\altaffiltext{6}{Department of Physics and Astronomy, Rutgers University, 136 Frelinghuysen Road, Piscataway, NJ 08854-8019.}
\altaffiltext{7}{Departamento de Astronom\'{\i}a, Universidad de Chile, Casilla 36-D, Santiago, Chile.}
\altaffiltext{8}{Sterrewacht Leiden, Leiden University, NL-2300 RA Leiden, Netherlands.}
\altaffiltext{9}{Kavli Institute for Astronomy and Astrophysics, Peking University, Beijing, China.}
\altaffiltext{10}{Space Telescope Science Institute, 3700 San Martin Drive, Baltimore, MD 21218, USA}

\begin{abstract}
We study the properties of a sample of 211 heavily-obscured Active Galactic Nucleus (AGN) candidates in the Extended
Chandra Deep Field-South selecting objects with $f_{24\mu m}$/$f_R$$>$1000 and $R$-$K$$>$4.5. Of these, 18 were detected in X-rays
and found to be obscured AGN with neutral hydrogen column densities of $\sim$10$^{23}$~cm$^{-2}$. In the X-ray undetected sample,
the following evidence suggests a large fraction of heavily-obscured (Compton Thick) AGN: (i) The stacked X-ray signal of the sample is
strong, with an observed ratio of soft to hard X-ray counts consistent with a population of $\sim$90\% heavily obscured 
AGN combined with 10\% star-forming galaxies. (ii) The X-ray to mid-IR ratios for these sources are significantly larger than that 
of star-forming galaxies and $\sim$2 orders of magnitude smaller than for the general AGN population, suggesting column densities
 of $N_H$$\gtrsim$5$\times$10$^{24}$~cm$^{-2}$. (iii) The Spitzer near- and mid-IR colors of these sources are consistent with 
those of the X-ray-detected sample if the effects of dust self-absorption are considered. Spectral fitting to the rest-frame
UV/optical light (dominated by the host galaxy) returns stellar masses of $\sim$10$^{11}$M$_\odot$ and $<$E(B-V)$>$=0.5, and reveals 
evidence for a significant young stellar population, indicating that these sources are experiencing considerable star-formation. This sample
of heavily-obscured AGN candidates implies a space density at $z$$\sim$2 of $\sim$10$^{-5}$Mpc$^{-3}$, finding a strong evolution in the 
number of  $L_X$$>$10$^{44}$~erg/s sources from $z$=1.5 to 2.5, possibly consistent with a short-lived heavily-obscured phase before an 
unobscured quasar is visible.
\end{abstract}


\section{Introduction}

Understanding the processes of galaxy formation and evolution is one of the outstanding problems of modern astronomy. It is now clear 
that the growth and feeding of the central black hole of an active galactic nucleus (AGN) and the formation of its host galaxy must
be intimately connected, as evidenced by the striking correlations between the properties of supermassive black holes and the stellar 
systems in which they reside \citep{magorrian98, gebhardt00, ferrarese01}. Furthermore, it is now strongly suspected that AGN activity is 
critical in the regulation of star formation in the host galaxy either by removing all the gas (e.g., \citealp{hopkins06,menci06}), by 
heating it (e.g. \citealp{croton06}) or both \citep{schawinski06}. However, an observational link between AGN activity and quenching of 
star formation has not yet been established (e.g., \citealp{schawinski09}). To understand how the AGN and galaxy formation processes 
are connected requires a detailed census of the AGN population as well as detailed study of individual objects.

The visibility and detectability of an AGN is determined by the interplay among black hole mass, accretion rate, gas distribution, dust 
geometry and the amount of star formation in the host galaxy. Thus, no single selection method provides a complete view of how galaxies and AGN
co-evolve. Many surveys (e.g., \citealp{barger01,cowie02}; see \citealp{brandt05} for a review) have focused on X-ray selection because 
X-ray emission is a universal feature of AGN accretion. However, X-ray selection has limited sensitivity to heavily obscured and/or
low accretion rate sources, even in the deepest surveys \citep{treister04}. In particular, examples of the most heavily-obscured AGN, 
the so-called Compton-Thick (CT) AGN, which have $\tau$$\sim$1 for Compton scattering or $N_H$$\sim$10$^{24}$~cm$^{-2}$, are mostly 
found in the local Universe.  Only a few of these sources have been identified at higher redshifts, although this is simply due to selection 
effects \citep{dellaceca08b}. In contrast, early AGN population synthesis models that can fit the X-ray background (XRB) spectral shape 
and intensity \citep{treister05b,gilli07} require $\sim$20-30\% CT AGN in order to explain the observed peak at $\sim$30~keV in the 
XRB radiation. Recently, \citet{treister09b} combined observations of local ($z$$<$0.05) AGN at very hard X-rays (E$>$20~keV), using the 
International Gamma-Ray Astrophysics Laboratory (INTEGRAL) and Swift satellites, with XRB models and concluded that the CT AGN 
fraction is only $\sim$10\% at these redshifts. They conclude that CT AGN represent a similarly small fraction of the total XRB radiation 
and total black hole growth. However, because only $\sim$2\% of the XRB comes from CT AGN at $z$$>$2 and only a few sources are 
known at such high redshifts, this population still remains basically unconstrained, with up to an order of magnitude uncertainty in number 
density \citep{treister09b}.

Because the energy absorbed at optical to X-ray wavelengths is later re-emitted in the mid-IR, it is expected that AGN, in particular the
most obscured ones, should be very bright mid-IR sources (\citealp{treister06a} and references therein), easily detectable in 
observations with the Spitzer observatory. Sources having mid-IR excesses, relative to their rest-frame optical and UV emission, have been
identified as potential CT AGN candidates at $z$$\sim$2 (\citealp{daddi07,fiore08,georgantopoulos08} and others). However, because of
the strong connection between vigorous star formation and AGN activity in the most luminous infrared sources \citep{sanders88}, the relative
contribution of these processes is still uncertain and remains controversial (e.g., \citealp{donley08,pope08}). 

In order to detect CT AGN up to high  redshifts, deep multiwavelength coverage is critical. The Extended Chandra Deep Field South (ECDF-S) is 
one of the best fields to carry out this study. The rich multiwavelength data available in the ECDF-S includes four $\sim$250~ksec Chandra pointings 
covering an area of $\sim$0.3~deg$^2$, and deep Spitzer data in the IRAC bands from the Spitzer IRAC/MUSYC Public Legacy in 
ECDF-S (SIMPLE) survey and at 24~$\mu$m from the Far-Infrared Deep Extragalactic Legacy Survey (FIDEL). The 24~$\mu$m depth in 
this field is $\sim$35~$\mu$Jy, comparable to the deep Great Observatories Origins Deep Survey (GOODS) observations.

In this paper, we present a study of the properties of the CT AGN candidates in the ECDF-S selected from their 24~$\mu$m to optical flux
ratio and rest-frame optical to UV colors. In Section 2 we describe the multiwavelength data used in this work, while our selection
scheme is presented in Section 3. The properties of the sources individually detected in X-rays are discussed in Section 4, while the remaining sources are studied
in Section 5. The near and mid IR colors, X-ray to mid-IR flux ratios and space density of our sources are presented in Section 6, and our
conclusions are reported in Section 7. We assume a $\Lambda$CDM cosmology with $h_0$=0.7, $\Omega_m$=0.3 
and $\Omega_\Lambda$=0.7, in agreement with the most recent cosmological observations \citep{spergel07}.

\section{Multiwavelength Data}

\subsection{X-ray Data}

The full ECDF-S was covered by Chandra as part of a Cycle 5 guest observer program (PI: N. Brandt). The total area covered is 
$\sim$0.3 deg$^2$ to a depth of $\simeq$230 ksec. Details about these observations, images and catalogs were presented separately 
by \citet{lehmer05} and \citet{virani06}. The work of \citet{lehmer05} reports the finding of 762 sources to flux limits of 
1.1$\times$10$^{-16}$ and 6.7$\times$10$^{-16}$~erg~cm$^{-2}$s$^{-1}$ in the soft (0.5--2 keV) and hard (2--8 keV) bands. The catalog 
of \citet{virani06} includes 651 sources to similar flux limits, using a more conservative rejection of periods of higher background, thus 
reducing the number of spurious sources but excluding a few real sources as well. In this work, we use mainly the \citet{virani06} catalog. However, 
when performing stacking analysis we further exclude sources individually detected in the \citet{lehmer05} catalog and/or in the deeper 
1 and 2 Msec observations available in the central $\simeq$0.1~deg$^2$, presented by \citet{alexander03} and \citet{luo08} respectively.

\subsection{Mid-IR Data}

The ECDF-S was observed extensively at near and mid-IR wavelengths by the Spitzer Space Telescope. The central 10$'$$\times$16$'$ region 
was covered using both the Infrared Array Camera (IRAC; \citealp{fazio04a}) and the Multiband Imaging Photometer for Spitzer (MIPS; \citealp{rieke04}). 
These observations were performed as part of the GOODS survey and guaranteed time programs. More details about these observations were presented 
by \citet{treister06a} and R. Chary et al. (2009, in prep.).

The extended 30$'$$\times$30$'$ region was observed by IRAC as part of the SIMPLE survey. The flux limits for 
the SIMPLE observations are 0.76, 0.4, 5.8, and 3.6 $\mu$Jy in the 3.6, 4.5 and 5.7, and 8 $\mu$m bands, respectively, or $\sim$3--5 times shallower 
than the GOODS observations. More details about the IRAC coverage of the ECDF-S were reported by \citet{cardamone08} and M. Damen et al. (2009, in prep.). 
When required, the conversion factors from flux density to Vega magnitudes provided by \citet{fazio04a} were assumed. The ECDF-S was also observed at longer wavelengths in the FIDEL survey, which obtained data at 24, 70 and 160~$\mu$m. Because of the significant decrease in sensitivity at longer wavelengths, mainly
due to source confusion, only the 24~$\mu$m data are considered in this work. Details about the data reduction and catalog creation were presented 
by \citet{treister09a}. The approximate flux limit of the 24~$\mu$m data used here is $\sim$35$\mu$Jy. For comparison, the flux limit of the 
GOODS 24~$\mu$m data is $\sim$12$\mu$Jy \citep{daddi07}, while \citet{fiore08} studied only sources with $f_{24}$$>$40$\mu$Jy in the Chandra Deep
Field South (CDF-S) proper. The work of \citet{fiore09} in the COSMOS field used only the shallow cycle 2 MIPS data to a flux limit of $\sim$550~$\mu$Jy.

\subsection{Optical/Near-IR Data}

The ECDF-S was observed at optical wavelengths using both ground-based and space telescopes. Deep images (to $V$$\sim$26.5) in the $UBVRI$ were 
obtained using the Wide Field Imager (WFI) on the 2.2m telescope at La Silla. These data were made public by the ESO Deep Public Survey 
\citep{mignano07}, COMBO-17 \citep{wolf04}, and Garching-Bonn Deep Survey (GaBODS; \citealp{hildebrandt06}) teams. Imaging to $z'$=23.6 AB, 
was performed using the MOSAIC-II camera mounted on the Blanco 4m telescope at Cerro Tololo \citep{gawiser06b}. The Hubble Space Telescope
observed the ECDF-S in the $V$ and $z'$ bands as part of the Galaxy Evolution from Morphology and SEDs (GEMS) survey \citep{rix04}. Deep near-IR 
coverage was obtained using the CTIO 4m telescope with the Infrared Sideport Imager (ISPI), reaching a magnitude limit of $\sim$22 (AB) 
in the $JHK_s$ bands \citep{taylor09}, as part of the Multiwavelength Survey by Yale-Chile (MUSYC; \citealp{gawiser06a,treister07}).

In order to obtain high-quality photometric redshifts, the ECDF-S was imaged in 18 medium-band optical filters using the wide-field Suprime
Camera on the Subaru telescope. These filters range from 4270 to 8560\AA~ and are optimized for photometric redshift determinations \citep{hayashino00}. The 
data reach an average depth of $\sim$26 magnitudes (5$\sigma$, AB), extending redshift determinations to sources much fainter than most spectroscopic 
redshift surveys. More details about these observations, data reduction and catalogs will be presented by C. Cardamone et al. (2009, in prep.) and in
Section~\ref{photo_z}.

\subsection{Spectroscopic Redshifts}
\label{spec_z}

Rich optical spectroscopic data exist in the ECDF-S. We took advantage of the ESO/GOODS-CDFS spectroscopy master 
catalogue\footnote{Available at http://www.eso.org/science/goods/spectroscopy/CDFS\_Mastercat/} which compiles spectroscopic data from 11 
different sources, mostly using 8-10 meter class telescopes. In particular, many of the spectroscopic redshifts for the X-ray sources in the CDF-S proper 
were reported by \citet{szokoly04} using the VLT telescopes with the FORS1/FORS2 spectrographs. In addition, spectroscopic observations for 339 
X-ray sources in the ECDF-S field using the VLT/VIMOS and Magellan/IMACS are reported by \citet{treister09a}.

Given the very faint optical magnitudes of the ECDF-S X-ray sources, obtaining redshifts and identifications is a very hard, time-demanding task. In particular,
optical spectroscopy is often out of reach, even for state-of-the-art 10m-class telescopes. However, near-IR spectroscopy offers a higher
chance of success, not only because these sources are significantly brighter at these wavelengths, but more importantly because, based on the experience
with similar sources in other fields, they are found at redshifts $z$=1--3, where the typical optical emission lines are shifted to the
near-IR. For the brighter sources we used the Spectrograph for INtegral Field Observations in the Near Infrared (SINFONI; \citealp{eisenhauer03,bonnet04}) at the 
VLT, and the Multi-Object InfraRed Camera and Spectrograph (MOIRCS; \citealp{ichikawa06,suzuki08}) at the Subaru 8m telescope. For the remaining
sources in our sample we have to rely on photometric redshifts, which are relatively accurate because of the deep Subaru medium-band photometry described
below.

\subsubsection{SINFONI Data}

A total of 16 hours were granted to observe four X-ray sources in the ECDF-S with $V$-$K$$>$7 using the VLT/SINFONI Integral Field Unit (IFU) as part
of program 080.A-0612. The observations were carried out in service mode by the Paranal staff in the October 2007- January 2008 period. We attempted
to observe one of the sources (XID 647) with the Laser Guide Star Facility; however, due to technical problems, these observations were cancelled. Of 
the remaining three sources, two have IR-red excesses, XIDs 277 and 580, and one, XID 57, was not detected at 24~$\mu$m despite a very red 
optical/near-IR color ($R$-$K$=6.56),  and hence is not included in the sample discussed in this paper.

For these sources we used SINFONI in the 0.25$''$ spatial resolution mode, which provides a field of view of 8$''$$\times$8$''$. The H+K grism, which provides 
a wavelength coverage of 1.45--2.45~$\mu$m and a resolution of R$\sim$1500, was used. The total exposure time was 3 hours for XID 580 and 2.7 hours for 
XID 277. In each case, a square dithering pattern with 4$''$ offsets was performed so that the source was always in the field of view. This pattern was repeated 
four times. 

Data were reduced using the official ESO SINFONI pipeline v1.9.4\footnote{The SINFONI pipeline can be found at http://www.eso.org/sci/data-processing/software/pipelines/sinfoni/sinfo-pipe-recipes.html}. The SINFONI data reduction cookbook\footnote{http://www.eso.org/sci/facilities/paranal/instruments/sinfoni/doc/}, provided by the Paranal Science Operations group was followed. Briefly, raw data were first cleaned to remove bad lines created by the presence of hot pixels in the bias region. Then data are
bias-removed and flat-fielded. Wavelength calibration was performed using arc lamps. Distortions were corrected using the provided wavelength
maps. Finally, a data-cube was reconstructed using the sinfo\_rec\_jitter routine. Flux calibration and removal of telluric features was performed using a standard
star, typically a solar analog, observed within 2 hours of each science observation. Cube visualization and extraction of a 2-dimensional pseudo-slit was
done using the QFitsView tool\footnote{Available at http://www.mpe.mpg.de/$\sim$ott/QFitsView/}. Extraction of the 1-dimensional spectrum, smoothing and flux 
calibration were performed using standard IRAF tools.

\subsubsection{MOIRCS Data}

Source XID 322 was observed using the MOIRCS near-IR multi-object spectrograph mounted at the Subaru telescope
 on the night of December 18, 2008. MOIRCS was used in its multi-object mask mode, in order to use a sample of nearby bright stars to do the
 source acquisition. The $HK500$ grism, which provides a resolution of $\sim$700 in the 1.3--2.5~$\mu$m wavelength range, was used. 
Ten-minute individual integrations were performed, after which the source was dithered by 3$''$ along the slit following a standard ABBA routine 
in order to allow for an accurate sky subtraction. The total exposure time was 3 hours. Data reduction was done using the IDL-based MOIRCSMOSRED 
package created by Youichi Ohyama. Basic steps include extraction of the slit spectrum, sky-subtraction, flat fielding, flexure correction, co-adding of
individual AB spectra to finally combine the resulting sky-subtracted spectra. Wavelength calibration was performed using OH sky-lines as reference and
flux calibration was based on observations of HD20758, a G2V star, taken during the same night. 

\subsection{Photometric Redshifts}
\label{photo_z}

In order to obtain highly accurate photometric redshifts for all of the optically detected sources in the region of the ECDF-S, we obtained 18 medium band 
images of the field. Our deep medium-band Subaru imaging reaches $R_{AB} \sim 26$, 2 magnitudes deeper than previous CDFS medium band 
imaging (COMBO-17 reaches $R_{AB}\sim24$ with 12 medium band filters and 10\% accuracy; \citealp{wolf04}), and provides accurate 
redshifts ($\sim 1\%$ accuracy) and detailed spectral energy distributions for $\ge 90$\% of the detected X-ray sources. The Subaru images reach 
nearly all of the X-ray selected AGN. The greater number of filters and their even spacing give an effective spectral resolution 
of $R = \lambda / \Delta \lambda \sim 23$ \citep{taniguchi04}. The catalog was created using the deep BVR image for detection \citep{gawiser06a}, and 
contains roughly 60,000 sources with detailed spectral energy distribution. The catalog includes color-matched photometry from all 18 medium bands, in 
addition fluxes from the optical and near-IR ground based imaging (see \citealp{taylor09} for more details on the broad band data). Details of the 
medium-band data reduction and characteristics are provided in Cardamone et al. (2009, in prep).

In order to obtain accurate photometric redshifts, we used EAZY (Easy and Accurate Zphot from Yale), a program optimized to provide high quality redshifts 
over 0$<$$z$$<$4 \citep{brammer08}. EAZY is a new user friendly, full featured redshift fitting code, including a user-friedly interface based on 
HyperZ \citep{bolzonella00} and the use of priors (e.g., \citealp{benitez00}). The template set and the magnitude priors are based on semi-analytical models 
which are complete to very faint magnitudes rather than highly biased spectroscopic samples, and so are particularly useful for samples of objects such as 
our optically-faint mid-IR-selected sources, where complete spectroscopic calibration samples are not available. Further, EAZY introduces a template error 
function to account for wavelength-dependent template mismatch rather than relying on minizing the scatter between photometric and spectroscopic 
redshifts of the subsample of sources bright enough to have spectroscopic redshifts. EAZY allows for linear combinations of the templates sets. We make 
use of this feature including an additional broad-line AGN template when fitting the X-ray sources. EAZY assigns each object a redshift by marginalizing 
over the full posterior probability distribution rather than maximizing the likelihood.

We compute the photometric redshifts for the sources detected in the combined BVR image using the default EAZY parameters, adopting the default template 
error function and R-band photometric prior, and including the optical and near-IR ground based coverage from the MUSYC survey \citep{taylor09} in addition 
to our medium band data. We find a median $\Delta z$/($1+z$) of 0.01 for all sources with spectroscopic redshifts out to $z$$\sim$5. Limiting ourselves to the 
subsample of X-ray sources, we do only slightly worse, $\Delta z$/($1+z$)=0.011, but the fraction of outliers doubles to 20\%. Overall, for $\Delta$$z$/(1+$z$), this 
is a factor of three better than is currently done using broad band data \citep{brammer08}. We find good quality photometric redshifts 
($q_z$$<$ 1; \citealp{brammer08}) down to R-band magnitude of AB=26 for 60\% of the sample. For many of the remaining sources poor fits are obtained 
because the photometry is too uncertain, due to faintness of the sources, or intrinsic variability in the source over the time period of which the photometry 
was taken \citep{salvato09}. There are also cases where the intrinsic SED may not be matched by those provided in the template set or degeneracies in 
color-z space can result in multiple peaks in the redshift-probability distribution. If no restriction in the quality parameter, $q_z$, is used the statistical 
quality of the photometric redshifts degrades only slightly, to $\Delta$$z$/(1+$z$)=0.016 with an outlier fraction of 27\%. Overall, with medium band 
photometry we have achieved highly accurate photometric redshifts for the majority of the sources in the ECDF-S down to very faint magnitudes.

\section{Optical/Mid-IR Selection of CT AGN Candidates}
\label{opt_midir_sel}

X-ray observations have been very efficient in finding unobscured and moderately obscured AGN up to high redshifts and low luminosities (e.g., \citealp{brandt05}).
However, highly obscured, particularly Compton-Thick AGN, have been significantly excluded even from the deep Chandra and XMM-Newton surveys. Because 
most of the radiation absorbed at X-ray and UV wavelengths is then re-emitted in the mid- and far-IR, recent studies at these energies, mostly taking advantage 
of Spitzer observations, have been very successful in finding heavily-obscured AGN candidates missed by X-ray selection up to high 
redshifts \citep{daddi07,alexander08,donley08,fiore08}.

In particular, \citet{fiore08} presented a selection method based on the 24~$\mu$m to R band flux ratio and $R$-$K$ color; specifically, $f_{24}$/$f_{R}$$>$1000
and $R$-$K$$>$4.5 (Vega) . According to simulations based on stacking of the X-ray signal, they estimate the fraction of heavily-obscured AGN in this
sample to be greater than 80\% \citep{fiore08,fiore09}. Similarly, \citet{daddi07} found a large number of CT AGN candidates by selecting sources
which show a significant excess in the star formation rate measured from IR light, compared to the value derived in the ultraviolet. However, \citet{murphy09}
showed, using mid-IR spectroscopy and adding far-IR photometric data for the sources selected using the \citet{daddi07} method, that in only
$\sim$30\% of the cases the observed mid-IR excess was due to the presence of an AGN, while in the majority of the sources the presence of
strong aromatic PAH features  explained the discrepancy in the derived star formation rates.

\begin{figure}
\begin{center}
\plotone{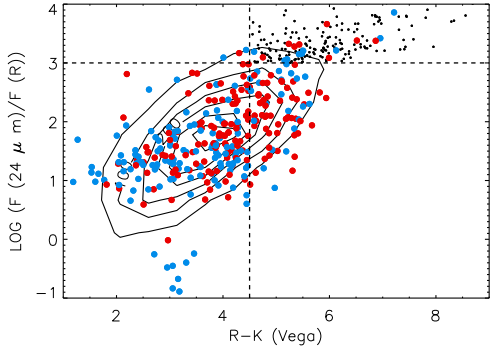}
\end{center}
\caption{Ratio of 24~$\mu$m to R-band flux density as a function of $R$-$K$ color. Contours show the location of 
all the MIPS-selected sources. {\it Red circles} show the location of X-ray sources with HR (define as (H-S)/(H+S) where S and H are
the background-subtracted counts in the soft and hard bands respectively) greater
than -0.3, while {\it blue circles} have HR$<$-0.3. {\it Small black circles} show the 193 sources in the IR-red excess region 
that were not detected individually in X-rays. While there is no clear trend of HR with optical ($R$-$K$) color, most of the X-ray hard
sources also have large $R$-$K$ colors.}
\label{f24R_RK}
\end{figure}

We applied the ``infrared-red'' selection criteria of \citet{fiore08} to the MIPS-selected sources in the ECDF-S to select CT AGN candidates, as shown 
in Fig.~\ref{f24R_RK}. Of the 7201 24$\mu$m sources detected in the FIDEL observations to a flux limit of $\sim$35~$\mu$Jy, 211 ($\sim$3\%) satisfy 
the $f_{24}$/$f_R$$>$1000 and $R$-$K$$>$4.5 cuts. Of the 651 X-ray detected sources in the \citet{virani06} catalog, 18 are found in this 
region, $\sim$2.8\%, a similar fraction as in the general population. The fraction of infrared-red sources with a direct X-ray detection is 
$\sim$9\%, smaller than the 16\% reported by \citet{fiore08} for the CDF-S observations, and significantly smaller than the fraction of direct X-ray 
detections in the COSMOS field (40\%; \citealp{fiore09}). As can be seen in Fig.~\ref{f24R_RK}, the typical value of $f_{24}$/$f_R$ for the ECDF-S 
sources is $\sim$50-100, while $<$$R$-$K$$>$$\sim$3.5, so the IR excesses are extreme, making them good candidates to be heavily obscured AGN.

\section{X-ray Detected AGN}

A total of 18 sources in the IR-red excess region were significantly detected individually in X-rays and included in the catalog of \citet{virani06}. 
In addition, six more sources were detected in the 2-Msec Chandra observations covering the central CDFS region \citep{luo08}; two of the latter sources 
were included in the \citet{lehmer05} catalog, but were not reported by \citet{virani06}. Four of these six sources were not included in the \citet{alexander03}
 catalog, indicating that they were very faint in X-rays and only significantly detected with the additional 1 Msec of data. 

For comparison, \citet{fiore08} report that the same number of IR-red excess sources, 18, were directly detected in X-rays and included in the 1 Msec Chandra 
catalog, with four more marginally detected. This can be expected by the combination of deeper X-ray data (by a factor of 4) and smaller area (by a factor
of $\sim$7). Also, an important factor is the difference in depth of the $K$ band image used in their analysis. While the MUSYC ECDF-S images reach a magnitude 
limit  of $K$$\simeq$20.2, the GOODS-MUSIC catalogs reach almost two magnitudes deeper \citep{grazian06}. 

\begin{figure}
\begin{center}
\plotone{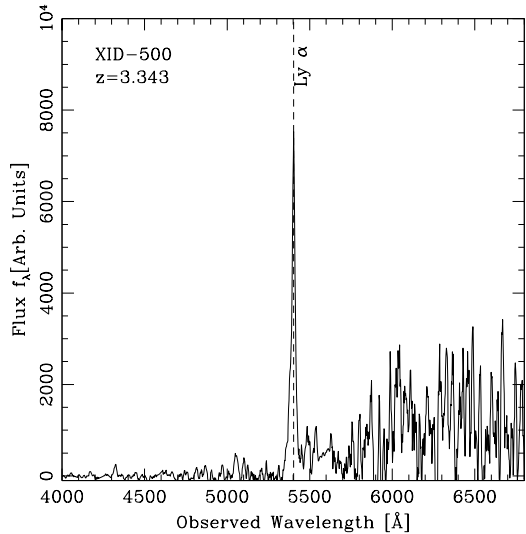}
\end{center}
\caption{Optical spectrum of X-ray source 500 obtained using VLT/VIMOS \citep{treister09a}. Wavelength is shown
in the observed frame. The redshift for this source is $z$=3.343, identifying the strong emission line at $\sim$5400 \AA~  as 
Ly$\alpha$, which is supported by the significant continuum decrement bluewards of this line. The photometric redshift for this 
source is 1.75$\pm$0.1 (99\% confidence level), but with a very bad $\chi^2$. The closest identification matching the phometric redshift 
and the observed emission line is [CIII]1909\AA~ at $z$=1.83; however, if this is the case, CIV at observed-frame 4383 \AA~ should also 
be visible. Hence, most likely the photometric redshift is wrong in this case.}
\label{spec_1982}
\end{figure}

After an extensive archival search, we found optical spectroscopy for two sources in our X-ray-detected sample, namely XIDs 480
and 500. Source XID480 is an intriguing source. Its redshift, $z$=1.603, was securely measured by \citet{szokoly04} based on deep VLT/FORS1-FORS2
observations. These authors classified XID 480 as a QSO-1, based on the soft X-ray spectrum. However, only narrow emission lines are present
in the optical spectrum. From X-ray spectral fitting \citep{treister09a}, we found that $N_H$=2$\times$10$^{22}$cm$^{-2}$, consistent
with the observed hardness ratio. Hence, this source is at the unobscured/obscured AGN boundary, which explains the discrepancy between the observed
soft X-ray spectrum and the red optical/near-IR colors. Source XID 500 was observed by VLT/VIMOS as part of our MUSYC identification program \citep{treister09a}. 
The optical spectrum of this source is shown in Fig.~\ref{spec_1982}. The spectroscopic redshift for this source is 3.343, based on a 
strong narrow emission line at 5402 \AA~ identified as Ly$\alpha$. In support of this interpretation, a strong decrement in the continuum is observed blueward
 of this line, while the continuum is clearly detected on the red side.

\begin{figure}
\begin{center}
\plotone{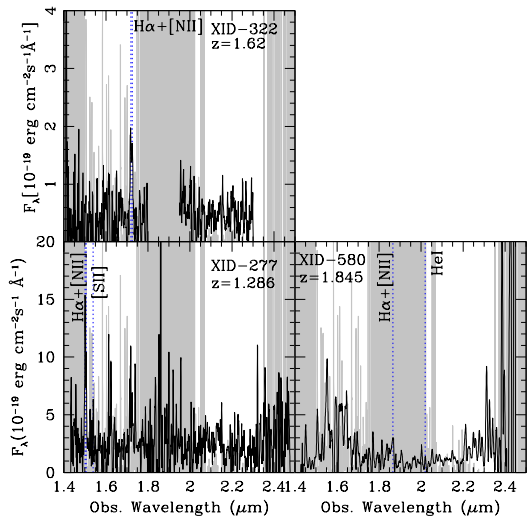}
\end{center}
\caption{Near-IR spectra for three X-ray-detected IR-red excess sources in the ECDF-S. Wavelength is shown in the observed frame. Shaded wavelengths show
spectral regions affected by low atmospheric transmission and OH sky lines \citep{rousselot00}. These spectra were obtained from VLT/SINFONI 
(XID-277 and XID-580) and Subaru/MOIRCS (XID-322). }
\label{spec_erxo}
\end{figure}

The near-IR spectra of sources 277 and 580 can be seen in Fig.~\ref{spec_erxo}. The redshift of XID 277 is secured by the presence of a strong emission line, 
which is marginally resolved at the SINFONI resolution and consistent with the wavelengths of $H\alpha$ and [NII] at a redshift of 1.286. In addition, a weak [SII] line
is present in the data. For XID 580, no obvious emission line is detected. However, on further inspection, two weak emission lines are found at wavelengths of
1.866 and 2.08 microns. These lines could be identified as $H\alpha$+[NII] and HeI respectively at a redshift of 1.845. The weak detection of these lines can be
explained by their location in a spectral region with very low atmospheric transmission. The main feature in the near-IR spectrum of XID 322 
is a strong emission line 1.722 microns. At the resolution of our setup, this line is barely resolved into two peaks, the largest one at 1.7179~$\mu$m and 
a smaller one at 1.7257 microns. This is consistent with $H\alpha$ and [NII] at a redshift of 1.621. No other feature is visible in this spectrum.

\begin{figure}
\begin{center}
\plottwo{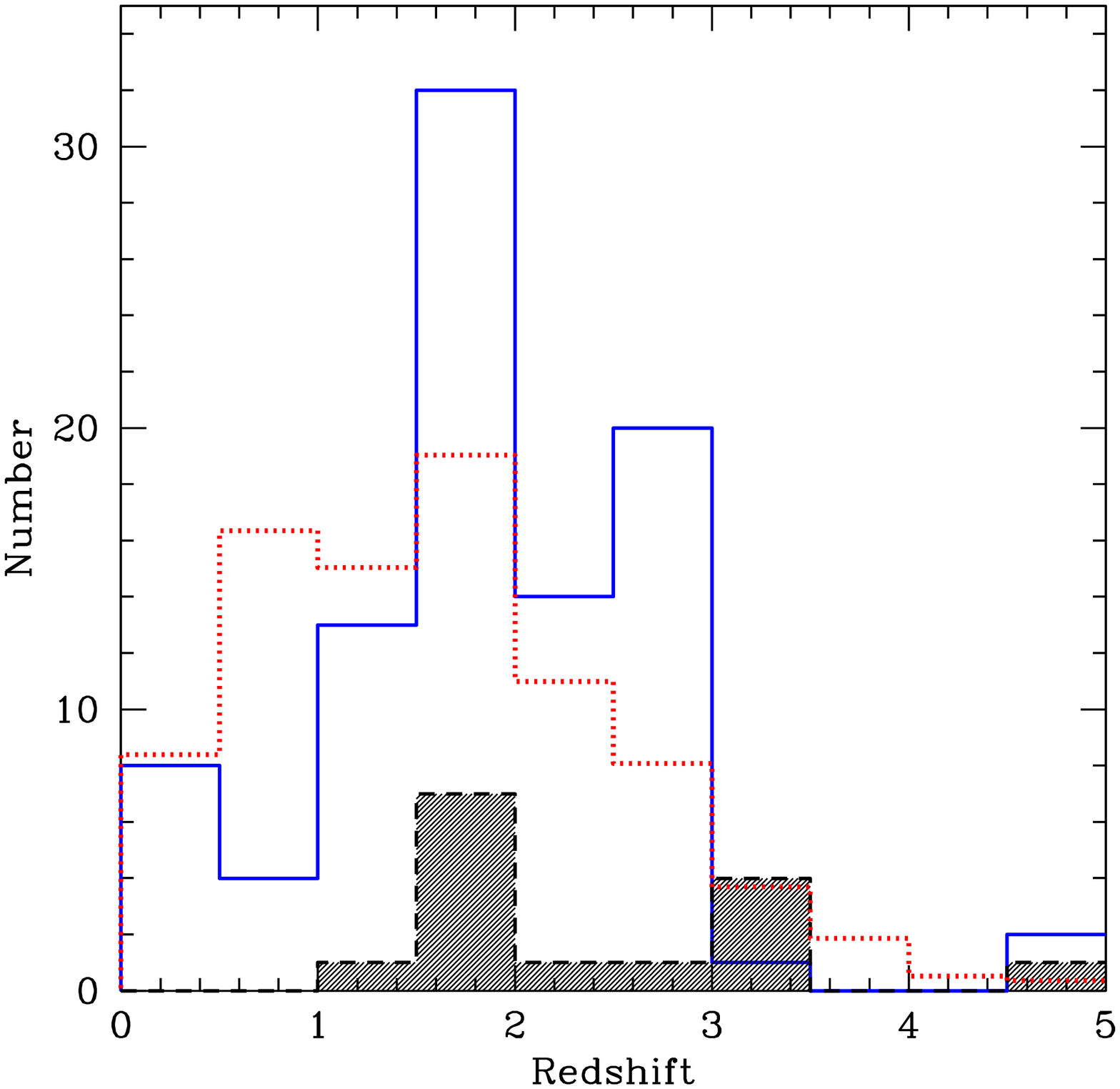}{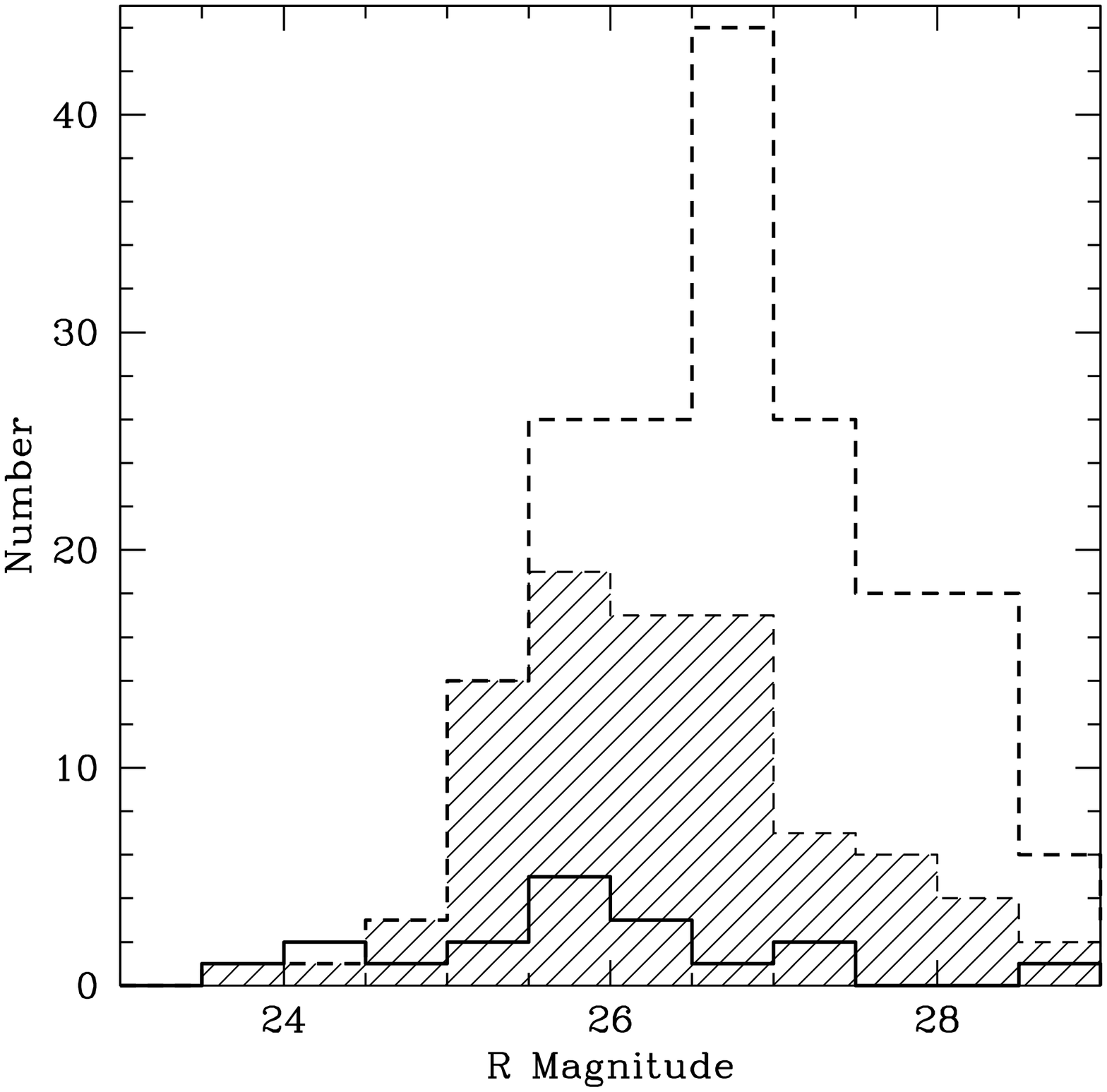}
\end{center}
\caption{{\it Left panel}: Redshift distribution for the 24 $\mu$m-selected sources in the ECDF-S. The {\it solid histogram} shows the distribution for 
the sources not detected in X-rays, while the {\it dashed hatched histogram} considers only the X-ray detected sources. A KS test shows that the 
hypothesis that these two distributions were drawn from the same parent distribution is refuted at the 84\% confidence level. The {\it dotted histogram} shows the 
slightly lower redshift distribution (divided by 1,000) for all the sources with a 24 $\mu$m detection and a measured photometric redshift in the 
ECDF-S. {\it Right panel}: distribution of R-band magnitude for the IR-red excess sources detected in X-rays ({\it solid histogram}), those not 
detected in X-rays ({\it dashed histogram}) and all sources with a measured photometric redshift ({\it hatched histogram}). Not surprisingly, spectroscopic 
and photometric redshifts are available preferentially for the optically-brightest sources. In addition, X-ray detected sources are brighter in the optical 
than the non-X-ray detected ones.}
\label{red_dist}
\end{figure}

The redshift distribution for the IR-red excess sources detected in X-rays,  including both photometric and spectroscopic measurements, is shown in
Fig.~\ref{red_dist}. The average redshift for our sample is 2.37 (median=1.85), larger than the value of 1.55$\pm$0.53 found by \citet{fiore09} for similar sources in the
COSMOS field. This is most likely due to the much brighter 24~$\mu$m flux limit of their sample. Indeed, \citet{fiore08} reported an average redshift of
2.1 in the deeper CDFS observations. Similarly, \citet{georgantopoulos08} estimated an average redshift of $\sim$2 for similar sources in the Chandra
Deep Field North. For our sample, the minimum redshift is 1.286 (XID 277) and the maximum is 4.65 (XID 303), although the latter is only a photometric
redshift.

\subsection{X-ray Properties}
\label{x_det_prop}

\begin{figure}
\begin{center}
\plotone{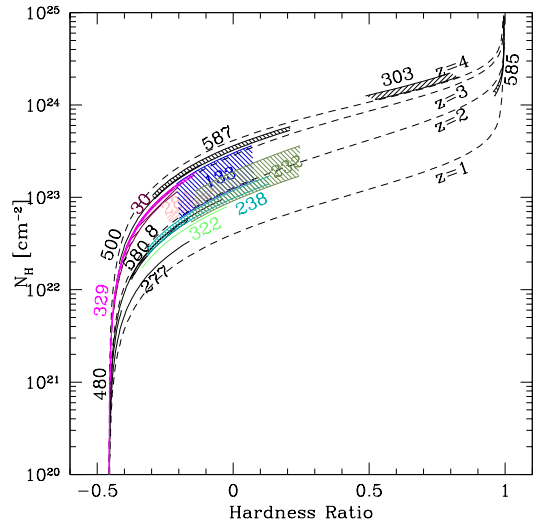}
\end{center}
\caption{Neutral hydrogen column density (N$_H$) as a function of hardness ratio for the 15 X-ray sources with either spectroscopic or photometric
redshifts. $N_H$ was calculated from spectral fitting for the X-ray brightest sources and from the HR for the remaining ones. For the two brightest sources, XID 480
and 284 we measured $\Gamma$=2.5$\pm$0.6 and 1.7$\pm$0.7 respectively. For the remaining sources, we assumed an intrinsic power-law spectrum 
with $\Gamma$=1.9 at the observed redshift of the source and correcting for galactic absorption. In general, these are very obscured sources in 
X-rays, with $N_H$$>$10$^{22}$~cm$^{-2}$, and two have Compton-thick absorption levels (XID 303 and 585).}
\label{hr_nh}
\end{figure}

Automated X-ray spectral fitting was performed for the sources in the ECDF-S using the Yaxx\footnote{Available at http://cxc.harvard.edu/contrib/yaxx/} software. 
An absorbed power-law was assumed for all sources. For the two brightest sources, XID 480 and 284, with more than 200 background-subtracted 
counts, the spectral slope and amount  of absorption were fitted simultaneously. In both cases, the observed spectral slope is consistent with the fiducial 
$\Gamma$=1.9 value \citep{nandra97}. For fainter sources, the spectral slope was fixed to $\Gamma$=1.9 and only the amount of absorption was fitted. For 
sources in which the number of X-ray counts is even smaller ($<$80 counts), it is not possible to perform spectral fitting. In those cases, we use the hardness
ratio (HR), defined as the ratio between the difference in counts in the hard and soft counts and their sum; the amount of absorption can be estimated 
from the HR by comparing the observed value with the expected counts for an intrinsic obscured power-law with $\Gamma$=1.9 at the redshift of the source 
and folding in the response of Chandra/ACIS. Because the same intrinsic spectral slope is assumed by the HR and spectral fitting methods, consistent results
are obtained. The observed HR values and the corresponding calculated $N_H$ for our X-ray detected sample are shown in Fig.~\ref{hr_nh}.

From our derived $N_H$ values we found two sources (13\%) with $N_H$$<$10$^{22}$~cm$^{-2}$, i.e., unobscured AGN, 11 (73\%) with 
10$^{22}$$<$$N_H$$<$10$^{24}$~cm$^{-2}$ (moderately obscured sources), and two Compton-thick AGN. The ratio of obscured to unobscured AGN
in this sample, 13:2,  is significantly higher than the average value of $\sim$3-4:1 found in deep surveys with no color cut (e.g., \citealp{treister04}). Furthermore, 
all the sources in this sample have $L_X$$>$$10^{43}$~erg~s$^{-1}$. Due to the strong dependence of the fraction of obscured AGN on 
luminosity (e.g., \citealp{ueda03,steffen03,barger05}), the fraction of obscured to unobscured sources is expected to be significantly 
lower ($\sim$1:2 at this luminosity; \citealp{treister09a}), making the excess in the relative number of obscured AGN more significant. This  
confirms that the IR-red excess selection preferentially find obscured sources, although only a small fraction of this sample reaches 
Compton-thick absorption levels.

\section{X-Ray Undetected CT AGN Candidates}

As was mentioned in Section \ref{opt_midir_sel}, a total of 211 24-$\mu$m sources in the ECDF-S were found in the IR-red excess region,
18 of them individually detected in X-rays and included in the catalog of \citet{virani06}. Of the remaining sources, eight were detected in X-rays
either by \citet{lehmer05} or in the deeper catalogs of \citet{alexander03} and \citet{luo08}. Here we focus on the 185 X-ray undetected obscured-AGN 
candidates. According to \citet{fiore08} the vast majority of these sources should be heavily obscured, even Compton-thick, AGN, in which the combination 
of photoelectric absorption and Compton scattering explains the lack of X-ray detection. In order to test this hypothesis, we study the nature of these
sources at many wavelengths through X-ray stacking, optical/mid-IR spectral analysis and IRAC colors.

\subsection{Photometric Redshift Distribution}

Contrary to the situation for the X-ray detected sources, where four sources had a measured spectroscopic redshift, none of the 185 X-ray undetected 
sources have a spectroscopy redshift. Hence, redshifts for these sources will be based on photometric measurements only. Because of the faintness 
of the optical/near-IR counterparts, accurate photometric redshifts are available for only 90 (49\%) sources. The remaining 95 sources were not detected
in the deep BVR image and hence it was not possible to compute photometric redshifts. The redshift distribution of these sources is shown 
in Fig.~\ref{red_dist}.

The vast majority of the mid-IR selected sources have photometric redshifts greater than one, with a median of 1.9, a significantly smaller value than
for the X-ray detected sources. Performing a Kolmogorov-Smirnov (K-S) test to these two distributions we found that the hypothesis that they
were drawn from the same parent distribution is marginally rejected at a 84\% confidence level. This small discrepancy can be due to the relatively 
large fraction of X-ray sources at $z$$>$3, in contrast with the strong drop at that redshift for the X-ray undetected sample.  We also found that 
X-ray detected sources are in general brighter than the X-ray undetected ones in both the 24~$\mu$m and $R$ bands, as can 
be seen in Fig.~\ref{red_dist}. A K-S comparing the $R$-band magnitudes for the X-ray detected and undetected sources found that the 
hypothesis that they are drawn from the same parent distribution is rejected with $>$99\% confidence. This difference in optical magnitude 
explains why the X-ray detected sample extends to higher redshifts. Furthermore, as mentioned before, the redshift completeness for the IR-red excess
sources not detected in X-rays is 49\%, while for the X-ray detected sources is 83\%, thus supporting the idea that the optical brightness explains
the difference in redshift distribution.

Compared to the general distribution of optically-detected sources in the ECDF-S, the IR-red excess sources are found at significantly higher
redshifts. This may be an effect of the selection, namely the $f(24\mu m)$/$f(R)$$>$1000 requirement. Because for obscured AGN the rest-frame optical/UV 
light is dominated by the host galaxy (e.g., \citealp{treister05a} and references therein), the $f(24\mu m)$/$f(R)$ is essentially a measurement of the 
AGN-to-host galaxy luminosity ratio. Hence, the fact that the 24~$\mu$m-selected sources are found mainly at $z$$>$1 is a consequence of the available volume
for high-luminosity sources. Furthermore, these sources do not have local analogs since even the most luminous nearby star-forming galaxies cannot
reach such high $f_{24\mu m}$/$f_R$ ratios \citep{dey08}. The high-redshift dropoff at $z$$\sim$3 can be explained by the faintness of the optical 
counterparts at such redshifts, thus making the determination of photometric redshifts very hard.
 
\subsection{X-ray Stacking}
\label{x_stack}

Thanks to its very low background, Chandra data can be efficiently stacked in order to study the average X-ray properties of a sample of individually-undetected
sources, thus achieving an effective exposure time of many megaseconds (e.g., \citealp{brandt01b}). In order to perform X-ray stacking
we used both the web-based CSTACK code\footnote{http://cstack.ucsd.edu/} developed by T. Miyaji and our own scripts, which are based
on CIAO \citep{fruscione06} version 4.1. In all cases we found consistent results using these two codes. Stacking was done using the final unbinned mosaics
of ECDF-S Chandra observations, which were produced as described by \citet{virani06}. The stacking was performed in two bands independently: soft (0.5--2 keV)
and hard (2--8 keV).

\begin{figure}
\begin{center}
\plottwo{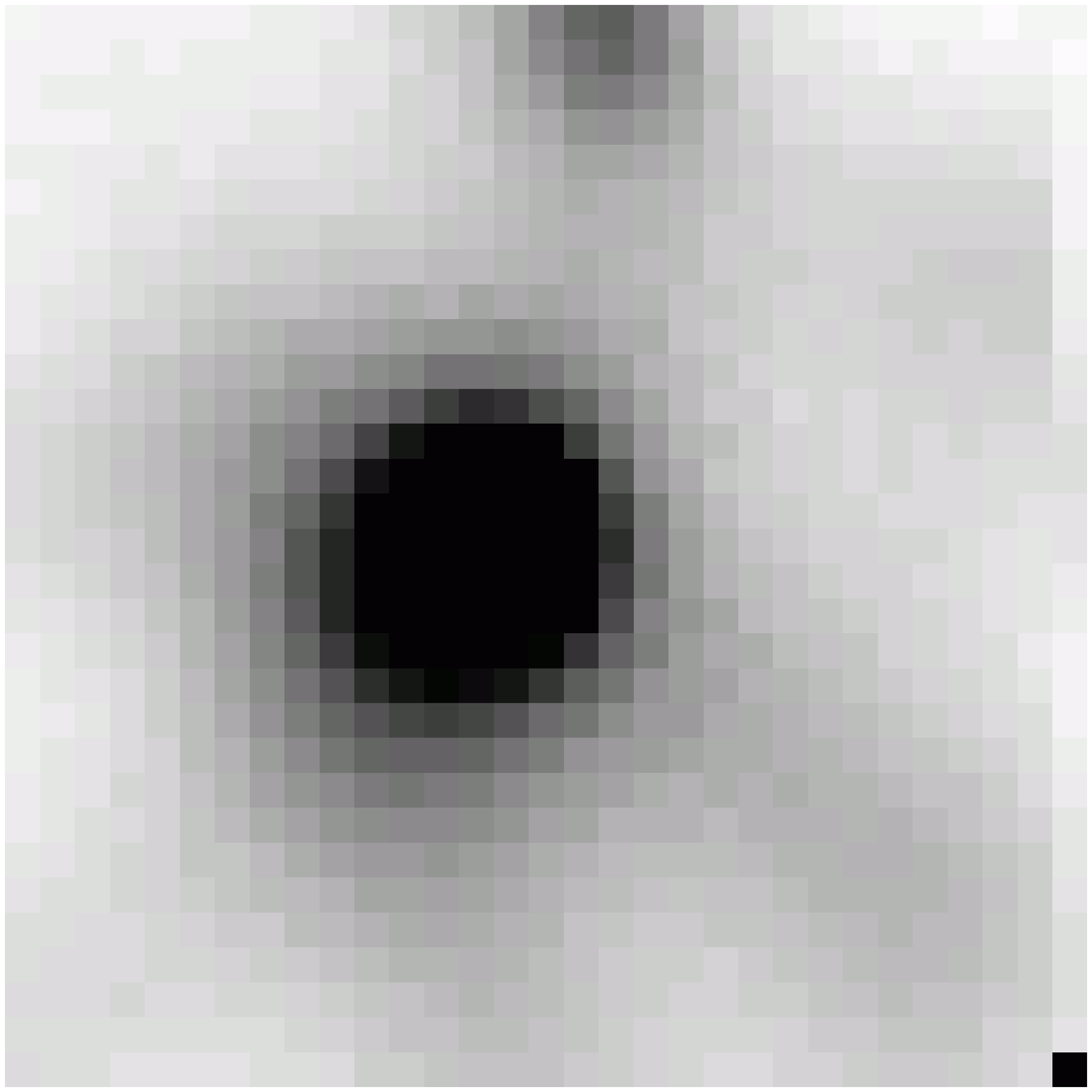}{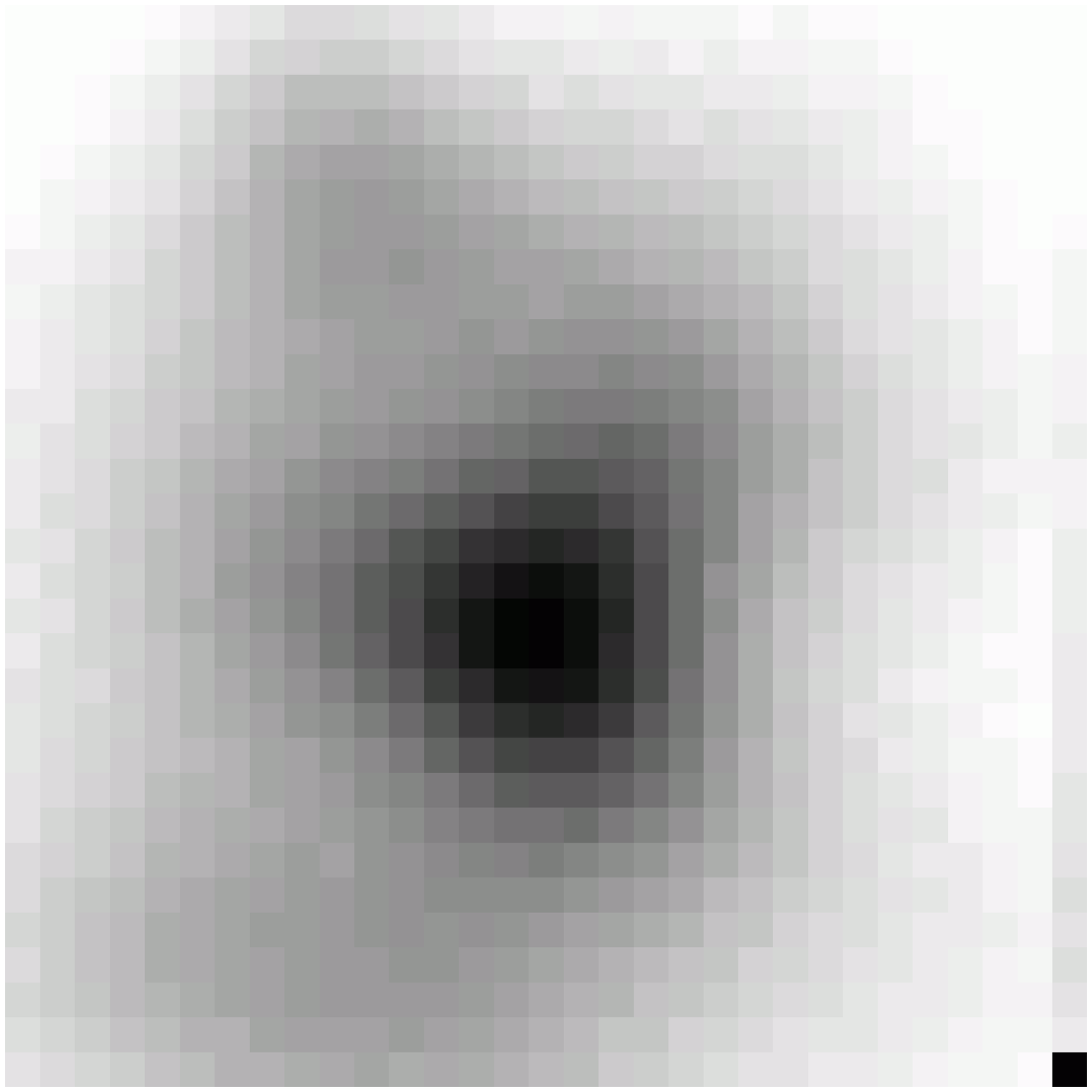}
\end{center}
\caption{X-ray stacking of 129 sources not detected individually by Chandra in the soft (0.5-2 keV; {\it left panel}) and hard (2-8 keV; {\it right panel})
bands. The equivalent exposure time is $\sim$30 Msec. A significant detection, $\sim$4-$\sigma$, is found in each band independently. The total 
background-subtracted counts are 68.1 in the soft band and 86.5 in the hard band, which correspond to average fluxes of $\sim$2 
and $\sim$8 $\times$10$^{-17}$erg~cm$^{-2}$s$^{-1}$ respectively.}
\label{stack_ecdfs_all}
\end{figure}

We started from the sample of 185 24~$\mu$m-selected sources in the ECDF-S not detected in X-rays. To avoid possible contamination and enable a more
accurate background measurement, we excluded 46 sources for which there was a detected X-ray source closer than 15$''$. After coadding the remaining
sources the effective exposure time was $\sim$30 Msec or $\sim$1 year. The stacked image in both the soft and hard bands is shown in Fig.~\ref{stack_ecdfs_all}.
Total source counts were measured in a 3$''$-radius circle. The background properties were estimated in two different ways: by using an annulus with a 7$''$ 
inner radius and 15$''$ outer radius and by performing photometry in random 3$''$ circles outside the central region. The results from both methods are 
roughly consistent with each other. We measured a total of 68 and 89 background-subtracted counts in the soft and hard bands. The background standard 
deviations are 14.1 and 32.8 counts respectively, which corresponds to a signal to noise of $\sim$4.8 in the soft band and $\sim$2.6 in the hard band. 
The significance of these detections was established by performing 500 independent Monte Carlo simulations, stacking 139 random positions in the
field, thus having the same effective exposure time and noise properties as the stacking of the 24~$\mu$m-selected sources. By studying the 
distribution of central background-subtracted counts in these simulations we found that, as expected, they are consistent with a Gaussian distribution
with standard deviations of 17 and 22 counts in the soft and hard bands respectively. This indicates that the signal detected for the 24-$\mu$m-selected
sources is significant at the $\sim$4-$\sigma$ level in each band. The latter measurement is more robust because it is based on 500 independent
determinations, compared to the twenty 3$''$ circles used in the method described previously, hence significantly reducing the effects of the 
faint undetected X-ray sources in the distribution of background counts.

In order to convert count rates into fluxes we used the Portable, Interactive Multi-Mission Simulator (PIMMS) tool\footnote{Web version available 
at http://heasarc.gsfc.nasa.gov/Tools/w3pimms.html}. Assuming the corresponding Chandra/ACIS response functions, an intrinsic power-law spectrum
with $\Gamma$=1.9, Galactic absorption of 8$\times$10$^{19}$cm$^{-2}$ \citep{kalberla05}, and intrinsic absorption of 10$^{24}$~cm$^{-2}$ at $z$=2,
we found conversion factors from counts/second to erg~cm$^{-2}$s$^{-1}$ of 6.5$\times$10$^{-12}$ in the soft band and 2.6$\times$10$^{-11}$ for the 
hard band. Hence, the average observed X-ray fluxes for the 24~$\mu$m-selected sources are $\sim$2.1$\times$10$^{-17}$~erg~cm$^{-2}$s$^{-1}$ in 
the soft band and $\sim$8$\times$10$^{-17}$~erg~cm$^{-2}$s$^{-1}$ in the hard band. If instead a lower column density, 2$\times$10$^{23}$~cm$^{-2}$, is 
assumed, the fluxes in the soft and hard bands are reduced by 17\% and 20\% respectively. Similarly, if a $z$=1 is assumed instead for a 
$N_H$=10$^{24}$cm$^{-2}$ spectrum the conversion for the hard band is increased by 27\%, while if $z$=3 is used the conversion is decreased
by 13\%. We can hence conclude that the uncertainties due to the assumption of an average spectral shape for the counts-to-flux conversion
generate uncertainties smaller than $\sim$30\%.

The HR for the stacked X-ray signal is 0.13$\pm$0.06. For comparison, for the sources individually detected in X-rays we found an average 
HR of -0.23$\pm$0.01, significantly softer than the value for the X-ray-undetected sample. This higher HR suggests the presence of
more obscured sources in the X-ray undetected sample. Converting directly this HR into a $N_H$ for a intrinsic power-law spectrum 
with $\Gamma$=1.9 and the average redshift of our sample, $z$$\simeq$2, we find that $N_H$$\simeq$1.8$\times$10$^{23}$~cm$^{-2}$. If
instead the extreme values of $z$=1 and $z$=3 are considered we obtain values of $N_H$=5$\times$10$^{22}$ and 
4$\times$10$^{23}$~cm$^{-2}$ respectively, i.e., a uncertainty of $\sim$2$\times$ can be expected due to the redshift distribution of our sources.
However, and more importantly, we note that the observed HR could be significantly affected by the presence of X-ray emission from 
non-AGN galaxies, which have in general a softer spectrum (e.g., \citealp{fabbiano89}). In order to estimate the fraction of obscured AGN relative to 
star-forming galaxies in a stacked sample, \citet{fiore09} performed simulations combining these two types of sources to match the observed HR. These 
simulations take into account the redshift distribution of our sources. Converting our observed HR into the bands used in their work (0.3--1.5 and 1.5-6~keV) we 
find that the fraction of heavily obscured AGN with $N_H$$>$10$^{23}$~cm$^{-2}$ is $\sim$90\%.

To understand the effects of the sources without either a photometric or spectroscopic redshift determination on our results we stacked separately
the 90 sources with a redshift measurement. A total of 25 and 31 background-subtracted counts were detected in the soft and hard band
respectively. The HR derived from these detections is 0.11$\pm$0.16, i.e., in very good agreement with the value found for the whole sample. This
indicates that while sources without a measured redshift can have a different redshift distribution, this does not affect our results significantly.
Similarly, splitting the sample with redshift measurements at $z$=2, roughly the median of the distribution, we detected 18 (soft band) and 27 (hard
band) background-subtracted counts for sources at $z$$<$2, while no significant detection was obtained for the $z$$>$2 sample. The HR
for the $z$$<$2 sample is 0.2$\pm$0.18, slightly larger but consistent with the value obtained for the whole sample.

\begin{figure}
\begin{center}
\plotone{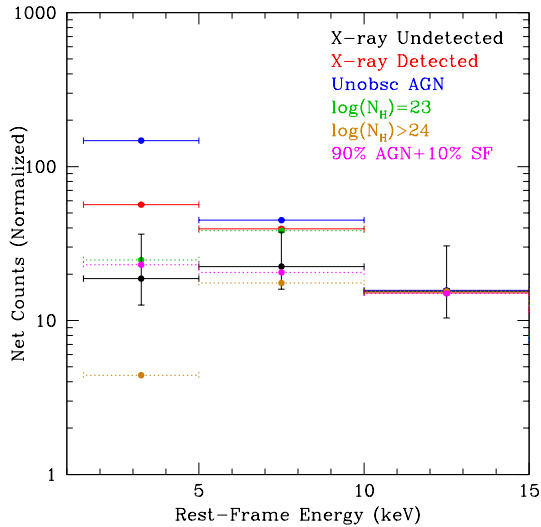}
\end{center}
\caption{Spectrum resulting from X-ray stacking performed in the approximate rest-frame. The {\it black points} show the background-subtracted counts
measured from the combination of 63 IR-red excess sources with measured redshifts that were individually undetected in X-rays, while the {\it red points} 
show the combined signal from the 15 X-ray detected IR-red excess sources. 
The {\it blue points} show the spectrum of source XID-312, a
luminous unobscured AGN in the ECDF-S at $z$=1.887. 
The last two spectra were normalized to the measurement for the X-ray undetected
sources at 10--15 keV, where the emission should be more isotropic. 
The {\it green} and {\it brown} data points show the expected spectral
shape for heavily obscured AGN with column densities of $N_H = 10^{23}$ and $>$10$^{24}$~cm$^{-2}$ (see text for details). The {\it magenta} data points
show the combined spectrum of a population of 90\% CT AGN and 10\% pure star-forming galaxies. The spectrum of the X-ray undetected sources can 
be well-explained by either an obscured AGN with $N_H$$\gtrsim$10$^{23}$cm$^{-2}$ or a mixed population of Compton-thick AGN and star-forming galaxies.
}
\label{rest_frame_stack}
\end{figure}

To further study the possible effects of the spread in redshift of our sample we also performed X-ray stacking in the approximate rest-frame, i.e., multiplying 
each observed photon energy by 1+$z$ before stacking. (This is not strictly correct since one should correct the incident photon energy rather than the 
observed photon energy. However, for the Chandra ACIS detectors, as for most X-ray detectors with modest energy resolution, a delta function of 
incident photons is redistributed over a broad range of energies and the response matrix cannot be inverted in a stable or unique way.) This pseudo-rest-frame 
stacking can be done only for the 90 sources with redshift measurements, which is roughly half the sample.
Excluding the sources closer than 15$"$ to a individually-detected X-ray source further reduces our sample to 63 sources. We perform
this X-ray stacking in four rest-frame bands: 1.5--5, 5--10, 10--15 and 15--24 keV. In the first three bands we obtained a signal-to-noise ratio
greater than $\sim$2, while there is no significant detection in the last band, most likely due to the rapid decrease in sensitivity of Chandra at high
energies and the reduced number of sources at high redshift in our sample. These results are shown in Fig.~\ref{rest_frame_stack}. In order to 
compare with sources with a known spectral shape, we also stacked the X-ray emission from the 15 IR-red excess sources individually 
detected in X-rays with a measured redshift. As described in section \ref{x_det_prop}, these sources are obscured AGN with an 
average $N_H$$\simeq$10$^{23}$~cm$^{-2}$. We also compare with the rest-frame X-ray spectrum of a luminous unobscured AGN found 
in the same field, XID 312, which according to \citet{treister09a} has a redshift of 1.887 and hence covers the same rest-frame energy range 
as our IR-red excess sample. As expected, the X-ray spectrum of the IR-red excess sources is significantly harder than those of the X-ray 
detected sources with similar optical/IR colors and a typical unobscured AGN.

To quantify the amount of absorption from the rest-frame X-ray stacked signal presented in Fig.~\ref{rest_frame_stack}, we compute 
the X-ray spectra of a heavily-obscured AGN with $N_H$=10$^{23}$  and $>$10$^{24}$ cm$^{-2}$. For the latter, we incorporate the effects
of Compton scattering, as described by \citet{matt99a}, and an Fe K$\alpha$ line at 6.4~keV with an equivalent width of 2~keV, typical of 
Compton-thick AGN \citep{gilli99}. 
Fig.~\ref{rest_frame_stack} shows that for the sample of 15 IR-red excess sources individually
detected in X-rays, an $N_H$ of 10$^{23}$cm$^{-2}$ explains the observed ratio of 5--10 to 10--15 keV flux; at lower energies, the contribution 
from less-obscured sources in this sample can explain the observed excess relative to the model predictions. In the case of the IR-red excess sources
individually undetected in X-rays, the spectrum in the 1.5--15 keV range can be well explained both by an obscured AGN with $N_H$$\gtrsim$10$^{23}$
cm$^{-2}$ or by a mixed population of a majority of Compton-thick AGN and star-forming galaxies, with the later providing most of the flux at E$<$5~keV.
For example, in Fig.~\ref{rest_frame_stack} we show the resulting X-ray spectrum of combining a population of 90\% CT AGN with 10\% pure
star-forming galaxies. This is fully consistent with our results derived from the X-ray signal stacked in the observed frame presented above. Hence, we 
conclude that the redshift spread does not affect our conclusions and therefore in this work we focus on the observed-frame measurements in
order to increase our sample and thus the significance of our results.
 
A possible dependence of the fraction of AGN on mid-IR flux (and thus luminosity) for X-ray selected sources was reported by \citet{treister06a} 
and \citet{brand06}. Hence, it is reasonable to expect a similar behavior for our 24 $\mu$m-selected sample. In order to test for this effect, we
separated our sample in three bins based on 24~$\mu$m flux, with the number of sources chosen to guarantee a signal-to-noise ratio for the 
X-ray stacked signal greater than two. The brighter bin includes 34 sources with 268$<$ $f_{24\mu m}$ ($\mu$Jy)$<$1292. The signal to noise
of the stacked signal in the hard band is $\sim$2.9. For these sources we measure a HR of 0.37$\pm$0.25, which using the simulations of \citet{fiore09} 
corresponds to a AGN fraction of $\sim$95\%. At intermediate 24~$\mu$m fluxes we stacked 33 sources with 155$<$$f_{24\mu m}$ ($\mu$Jy)$<$266. 
The detection of the stacked signal in the hard band is 2.1. In this case we found a HR of -0.02$\pm$0.01, or an AGN fraction of $\sim$80\%. Similarly, 
for the faintest 72 sources, with 53$<$$f_{24\mu m}$ ($\mu$Jy)$<$148 we found that HR=0.018$\pm$0.01, which also corresponds to an AGN 
fraction of $\sim$80\%, although in this case the hard-band detection is not significant ($\sigma$$<$2). In conclusion, although we find a weak trend 
of AGN fraction with 24~$\mu$m flux, it is not a statistically significant result.

\subsection{Optical/Mid-IR Spectral Fitting}
\label{sed_fit}

Because in obscured AGN the optical light is dominated by the host galaxy, it is feasible to study the stellar population in these sources. In order to do that, we 
fit the $UBVRIzJK$ spectral energy distribution of our IR-red excess sources to an ensemble of model star formation histories based on a 
two-burst scenario (see e.g. \citealp{ferreras02,schawinski07, kaviraj07}) using the stellar models of \citet{maraston05}. We allow the young burst 
to make up between 1\% and 100\% of the stellar mass and range freely in age up to the age of the old burst, which must be at least 500 Myr old. We 
vary the decay timescale of the young burst from 10 Myr (an instantaneous burst) to 5 Gyr (effectively constant SFR) and fix the metallicity of both 
components at the solar value. We apply a screen of dust using the extinction law of \citet{calzetti00} for starburst galaxies and allow the color 
excess E(B-V) to vary from 0 to 2.8. An AGN component was not added since for these sources the amount of obscuration of the nuclear region should be large
enough to completely absorb the UV and optical emission, in order to explain the red $R$-$K$ color. Only those sources with detections in sufficient bands
to have at least one degree of freedom (six bands) are fitted. For the majority of sources where we do not  have a spectroscopic redshift, we assume the 
photometric redshift to be the true redshift. We find that not all IR-red excess sources result in an acceptable value of the reduced $\chi^2$ statistic: of 
those sources with X-ray counterparts, 12 out of the 15 that have detections in at least six bands and a redshift measurement, have a 
reduced $\chi_{\rm r}^2 < 3.84$.

\begin{figure}
\begin{center}
\includegraphics[angle=90,width=0.45\textwidth]{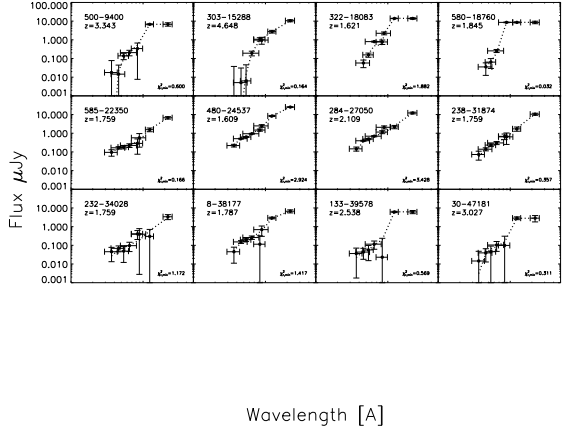}
\end{center}
\caption{Spectral energy distributions for the 12 IR-red excess sources in the ECDF-S individually detected in X-rays with measured spectroscopic or 
photometric redshift and detections in at least six bands. {\it Circles with error bars} show the observed photometric fluxes in the $UBVRIzJHK$  
bands. {\it Dotted lines} show the fit to the rest-frame optical light using galaxy templates. The median stellar mass for this sample 
is 5$\times$10$^{11}$~M$_{\odot}$, while the average extinction is E(B-V)=0.6.}
\label{seds_xrays}
\end{figure}

The nature of the available data limits what we can infer about the star formation history of these sources. We are sampling the rest-frame UV and blue 
optical part of the spectrum and expect a large amount of extinction in the UV. We also only have broad-band photometry, so as a result, we cannot 
constrain whether the current starburst that we do infer is an event in which most or all of the stellar mass is formed, or wether it is a less substantial 
burst with younger ages and higher amounts of extinction. In Fig.~\ref{seds_xrays} we show the resulting UV-optical spectral fitting for the X-ray detected 
sources. The median stellar mass for this sample is 4.6$\times$10$^{11}$~M$_{\odot}$, while the average extinction is E(B-V)=0.6. 

For our sample of IR-red excess sources not detected in X-rays, and requiring a redshift measurement and detection in at least six
bands, we found acceptable fits for 60 out of 84 sources. The most likely explanation for the poor fits are inaccurate photometric redshifts.
We marginalize over the nuisance parameters to investigate the stellar masses and levels of optical extinction inferred for our sample of IR-red 
excess sources. In general, they contain substantial stellar populations in an ongoing burst younger than 100 Myr and widely range in levels of optical 
extinction (0$<$E(B-V)$<$1)\footnote{This range in E(B-V) corresponds to 0$<$$A_{\rm V}$$<$4.05 assuming an $R_{\rm V}$ = 4.05 that 
the \citet{calzetti00} law uses.}. The best-fit stellar masses range between 10$^{9}$-10$^{12}$$M_{\odot}$ with a median stellar mass 
of $\sim$10$^{11}$$M_\odot$. We caution that the typical uncertainties in the individual stellar masses after marginalization are on the order 
of a factor of 10. We further investigate the effects of the uncertainty in the photometric redshift derivation in our derived stellar masses
and star-formation histories, by varying the redshift within its 1-sigma allowed range while doing spectral fitting. We found that the dispersion 
in the stellar masses is on the order of 0.1 dex, which is substantially smaller than the typical errors on individual masses. Furthermore, the typical
properties of the derived star-formation histories are not affected. Hence, we conclude that for the sources with good spectral fits, the errors on the 
derived stellar masses and star-formation histories are dominated by errors on the fluxes, in particular in the near-IR bands, and not by the photometric 
redshift errors.

\begin{figure}
\begin{center}
\includegraphics[angle=90,width=0.45\textwidth]{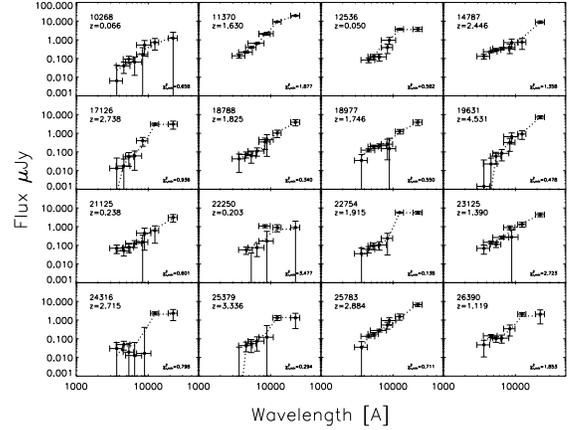}
\end{center}
\caption{Spectral energy distributions for a random sample of 16 IR-red excess sources not detected in X-rays individually. Symbols are the 
same as in Fig.~\ref{seds_xrays}. The mean stellar mass for the IR-red excess sources not detected in X-rays is 10$^{11}$M$_\odot$, while
the mean E(B-V) is 0.4. Both the stellar mass and extinction are slightly lower but consistent with the values found for the X-ray detected
sources.  In general we found evidence for a substantial young stellar population, indicating that the host galaxies of most of these sources
are experimenting significant moderately-obscured star formations episodes.}
\label{seds_nonxrays}
\end{figure}

\begin{figure}
\begin{center}
\includegraphics[angle=90,width=0.225\textwidth]{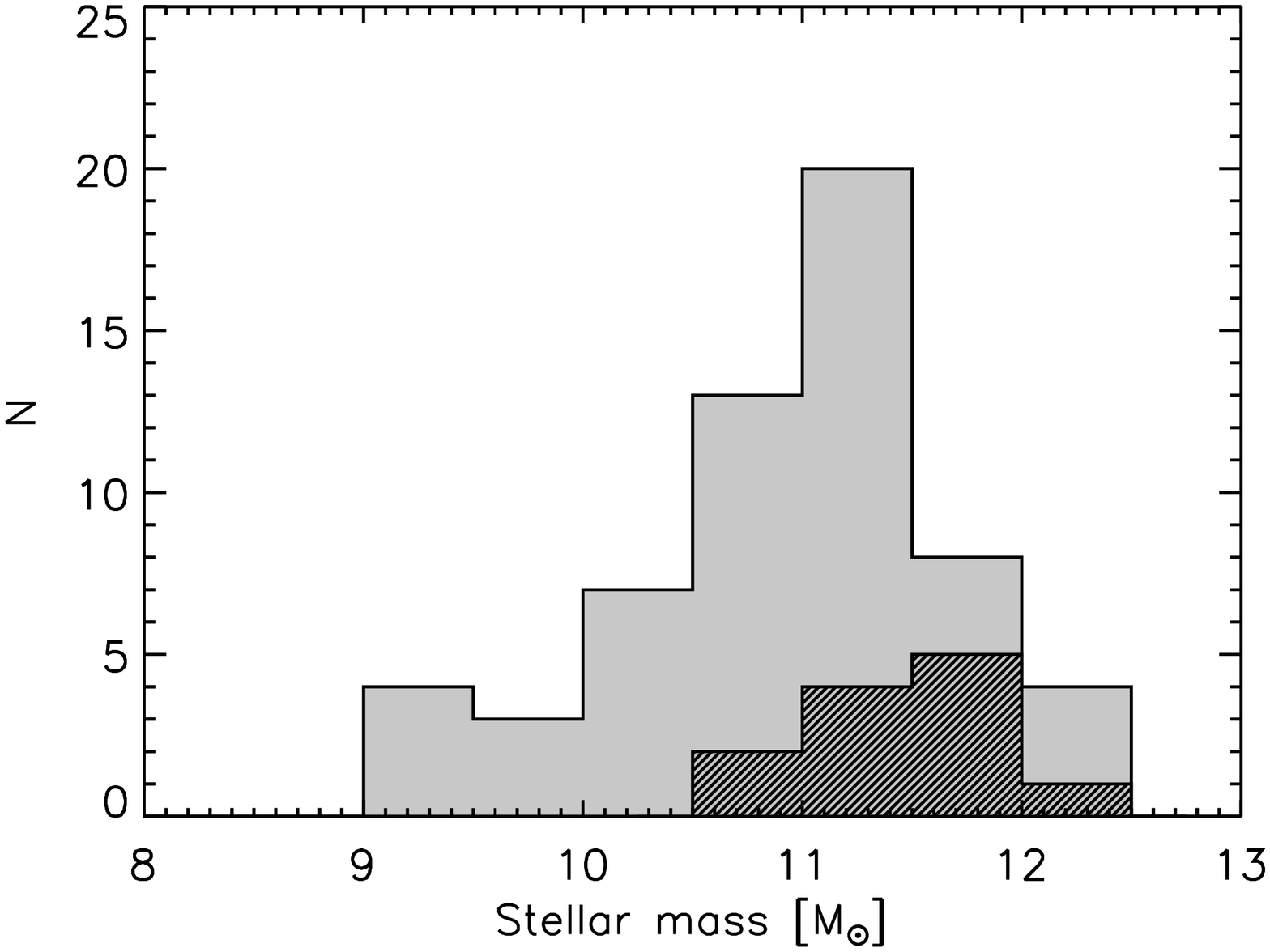}
\includegraphics[angle=90,width=0.225\textwidth]{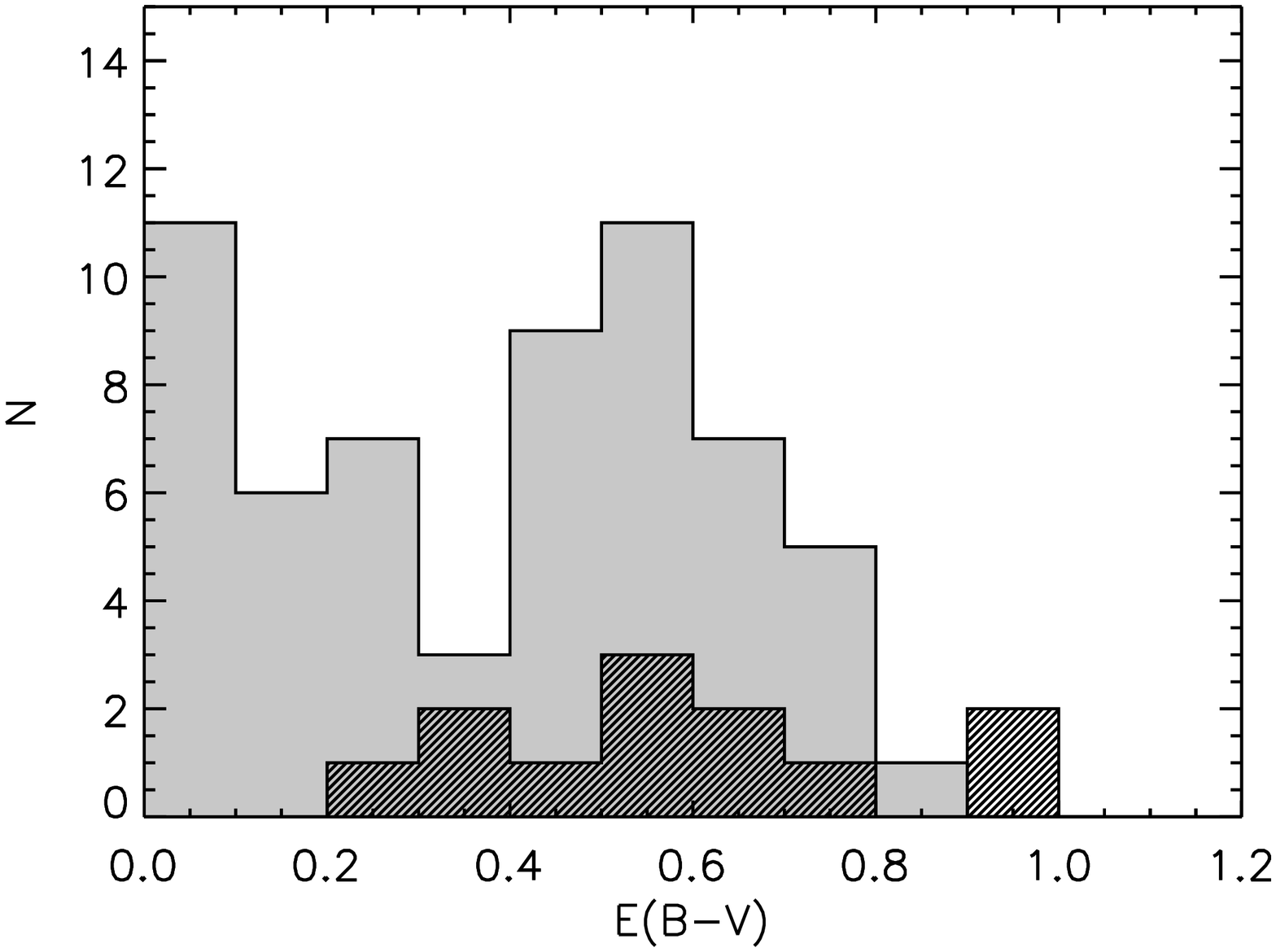}
\end{center}
\caption{{\it Left panel}: Distribution of stellar masses for the IR-red excess sources individually detected in X-rays ({\it hatched histogram}) and
not detected in X-rays ({\it solid histogram}). The median values and distribution for these two samples are very similar. Indeed, a K-S test 
performed to these distributions indicate that they are consistent with being drawn from the same parent distribution at the 90\% confidence
level.  {\it Right panel}: Distribution of extinction in the host galaxy, as measured from template fitting to the rest-frame UV/optical light. Symbols 
are the same as in the left panel. Contrary to X-ray absorption, which measures mostly the AGN obscuration, no significant differences are 
found between these two samples. In summary, we did not find major differences between the host galaxies of the X-ray detected and 
undetected sources.}
\label{sed_fit_props}
\end{figure}

In Fig.~\ref{seds_nonxrays} we show examples of spectral fitting for IR-red excess sources not detected in X-rays. We plot the histograms 
of the distributions of best-fit stellar masses and optical extinctions in Figure~\ref{sed_fit_props} - the hashed histogram in each panel 
represents the sources individually detected in X-rays, while the solid histogram represents those without detections. We perform a K-S to determine 
whether the E(B-V) and stellar mass distributions of the X-ray-detected and undetected IR-red excess sources are consistent with being drawn from 
the same parent population. We find that at the 90\% confidence level, both distributions are consistent, and therefore no significant differences are
found between the host galaxies of the X-ray detected and undetected sources.

\section{Discussion}
\label{discussion}

\subsection{Near and Mid-IR Colors}

As it was shown in Section \ref{x_stack}, the stacked X-ray signal for our 24~$\mu$m-selected sample indicates that most of these sources are
consistent with being heavily-obscured AGN. In addition to the X-ray spectral shape, the rest-frame near and mid-IR colors provide important 
clues about the obscuration levels in these sources and the importance of the AGN relative to the host galaxy. As expected from most 
dust re-emission models (e.g., \citealp{nenkova02}), the AGN spectrum at $\sim$3--10$\mu$m is significantly affected by the amount of material 
in the line-of-sight (due to the effects of self-absorption), which should be related to the obscuration of the X-ray emission. 

\begin{figure}
\begin{center}
\plotone{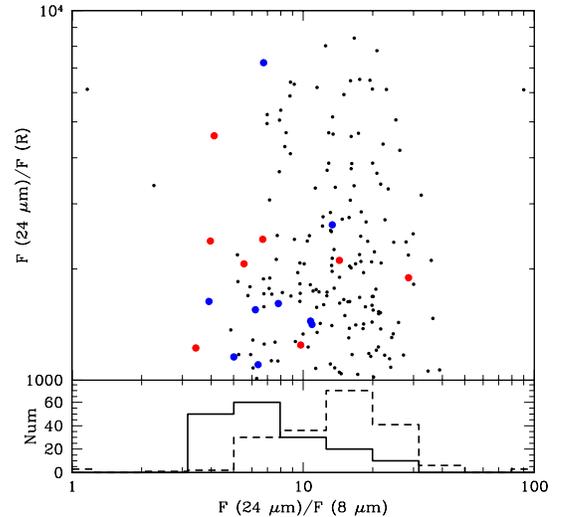}
\end{center}
\caption{{\it Upper panel}: Ratio of 24~$\mu$m to R-band flux density versus 24~$\mu$m to 8~$\mu$m flux ratio for
the 24~$\mu$m-selected AGN candidates in the ECDF-S. Same symbols as in Fig.~\ref{f24R_RK} are used. {\it Lower panel}: Distribution
of 24~$\mu$m to 8~$\mu$m flux ratio for the X-ray detected ({\it solid histogram}, increased by 10$\times$) and 
undetected ({\it dashed histogram}) sources. A clear difference between the two distributions is seen, with X-ray
detected sources having much bluer colors.}
\label{f24R_824}
\end{figure}

As can be seen in Fig.~\ref{f24R_824}, the distributions of $f_{24}$/$f_8$ flux ratios for X-ray detected and undetected sources are significantly
different. Indeed, performing a K-S test we found that the hypothesis that these distributions were drawn from the same parent distribution
is rejected at the $>$99.999\% confidence level. The distribution of $f_{24}$/$f_8$ ratios indicate a much bluer mid-IR spectrum for the X-ray
detected sources. The average values of $f_{24}$/$f_8$ are 8.7 for the X-ray sources and 24.9 for the X-ray undetected sources. At the typical 
redshift of our sources, these bands trace emission at rest-frame $\sim$8 and $\sim$3 microns. The dust re-emission models of
\citet{nenkova08b} suggest that these observed values could be explained by the same intrinsic torus parameters and two different viewing angles.
For example, for a torus model with N=10 (number of equatorial clouds), $\sigma$=30 (width of cloud's angular distribution), Y=30 (ratio of outer to inner
radii), q=2 (exponent of cloud's radial distribution assuming a power law) and $\tau_\nu$=40 (optical depth of an individual cloud; parameter 
definition given by \citealp{nenkova08a}), which are consistent with the models used by \citet{nenkova08b} to explain the mid-IR spectra of local 
Seyfert galaxies, the 8 to 3 micron flux ratios are consistent with the observed values for viewing angles of 30$^o$ for the X-ray detected sample 
and 90$^o$ (edge-on) for the X-ray undetected sources. These parameters are mentioned just as an example. By studying a total of 19,939 torus
models spanning a wide range of parameters we found that in $\sim$10\% of them the resulting $f_{24}$/$f_8$ ratios are consistent with the
observed values, taking into account the redshift distribution of our sources and assuming a viewing angle smaller than 30$^o$ for the X-ray detected
sources and greater than 70$^o$ for the undetected ones.

As shown in the sample of ultra-luminous IR galaxies presented by \citet{armus07}, in their Fig.~10, only a few sources reach values
of  the $f_{24}$/$f_8$ flux ratio as high as those observed in our sample of X-ray undetected IR-red excess sources. Furthermore, those sources
are mostly classified as LINERs, while using classification schemes based on optical emission lines (e.g., \citealp{kewley06}) they are found
in the ``composite'' region, i.e. they can be explain by a combination of an AGN and strong star-formation. Hence, we can conclude that
the observed $f_{24}$/$f_8$ ratios in our X-ray undetected sample strongly suggest the presence of a heavily obscured AGN in the majority
of them.

\begin{figure}
\begin{center}
\plottwo{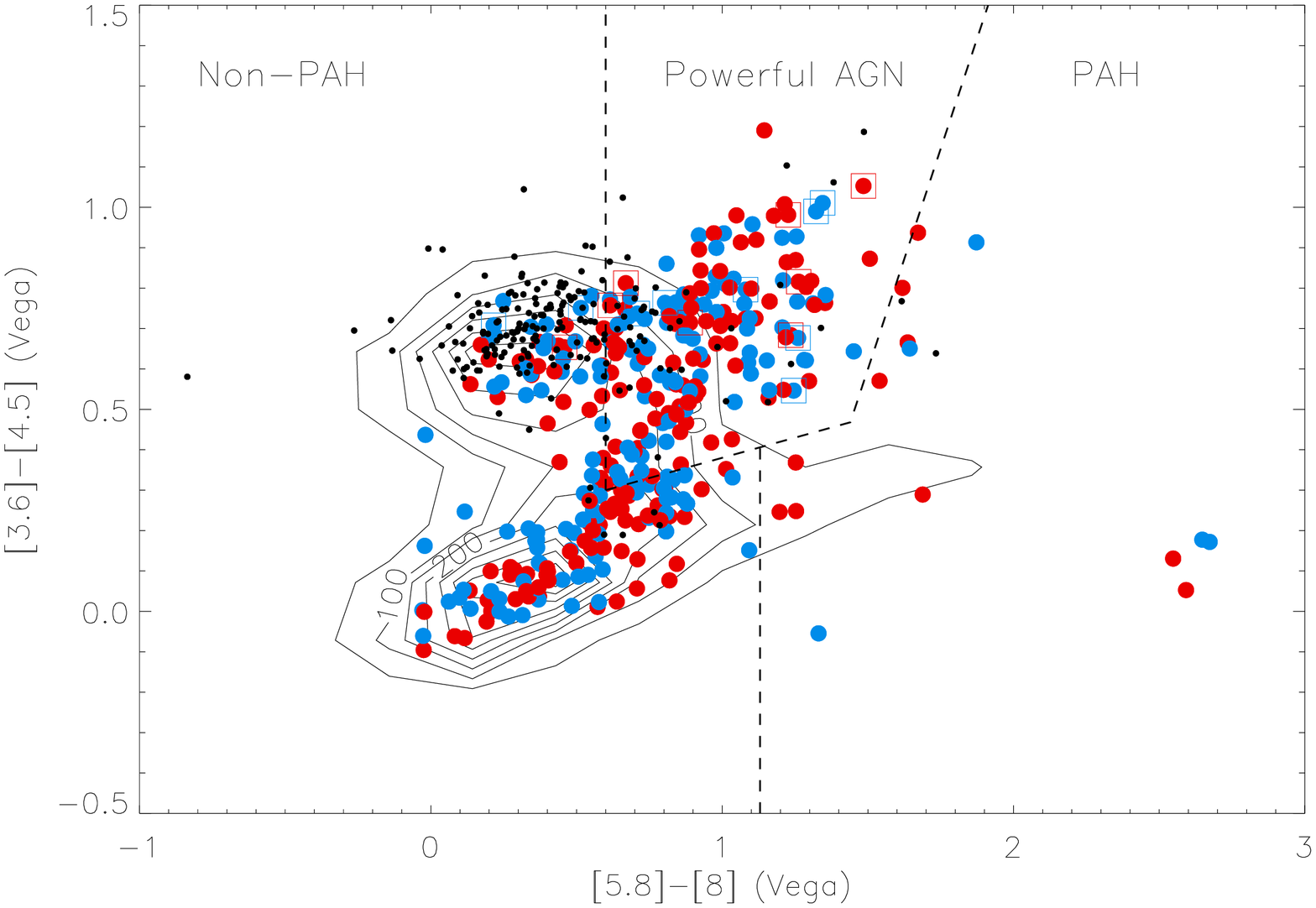}{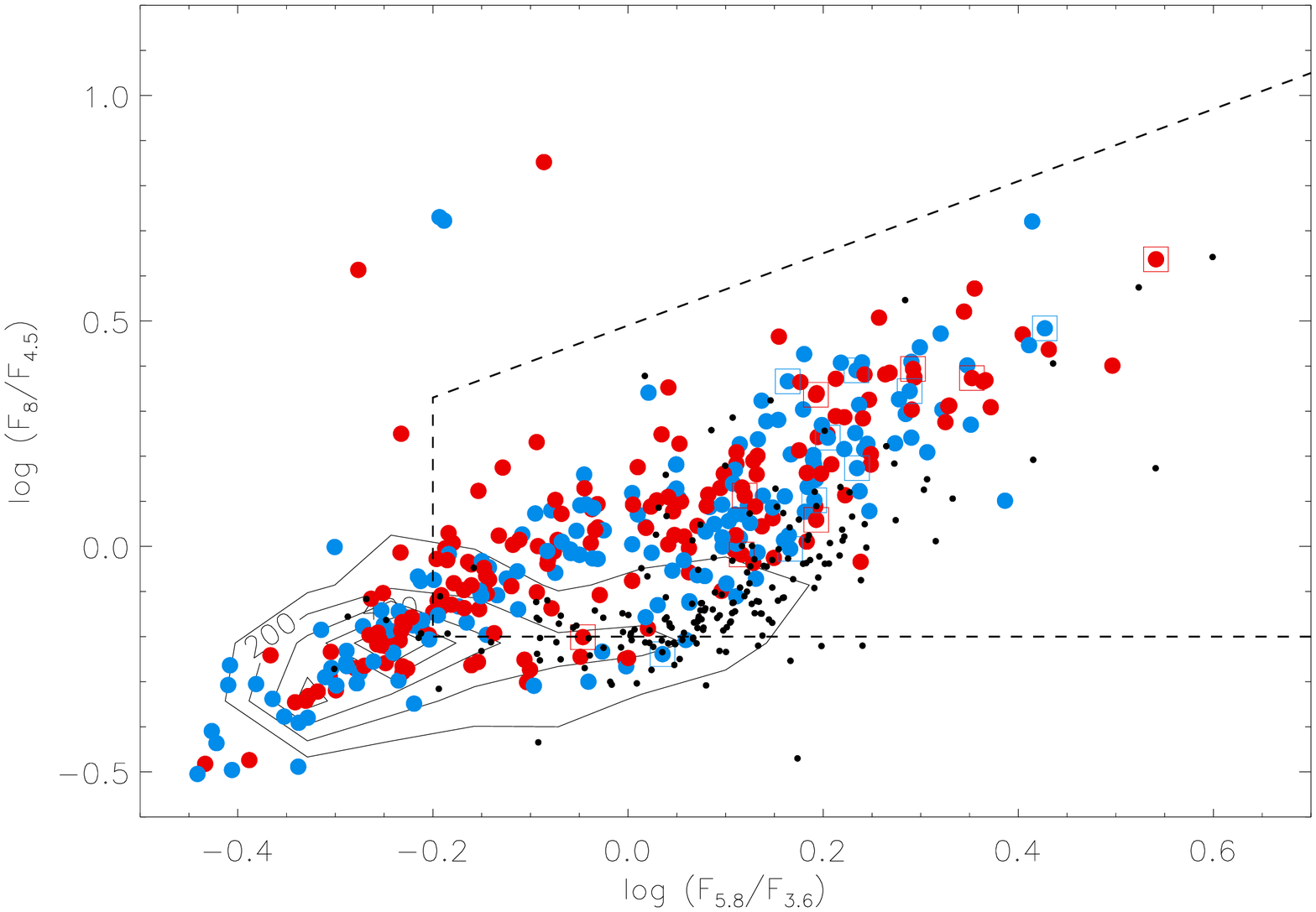}
\end{center}
\caption{Spitzer IRAC colors for the X-ray sources and obscured AGN candidates in the ECDF-S. Symbols are the same as in Fig,~\ref{f24R_RK}. {\it Left panel}:
[3.6]-[4.5] versus [5.8]-[8] colors. {\it Dashed lines} show the \citet{stern05} AGN region and the \citet{brand09} threshold for PAH-dominated sources. The 
IR-red excess sources not detected in X-rays are mostly outside the AGN region, mainly due to their relatively blue [5.8]-[8] color. {\it Right panel}: 
$f_8$/$f_{4.5}$ versus $f_{5.8}$/$f_{3.6}$ flux rations. The dashed region shows the AGN locus according to \citet{lacy04}. As can be seen in both panels, 
most of the X-ray detected 24~$\mu$m-selected sources are found in the AGN regions, suggesting that the rest-frame near-IR light is dominated by the 
nuclear emission. The $f_{8}$/$f_{4.5}$ flux ratio for the non-X-ray-detected sources is slightly lower than the one found in X-ray detected sources. Together 
with the difference found in Fig.~\ref{f24R_824} suggests that these sources  are more obscured versions of the X-ray detected AGN.}
\label{iracol}
\end{figure}

At the median redshift of our sample the Spitzer IRAC observations trace near-IR emission at $\sim$1 to $\sim$3 microns. This spectral region 
is dominated by a combination of stellar light, AGN continuum and hot dust emission. There have been several studies of
the IRAC properties of X-ray-selected AGN (e.g., \citealp{barmby06,cardamone08}). One of the main conclusions of the work of \citet{cardamone08} 
is that AGN show a large spread in near-IR colors and thus the IRAC color-color AGN selection methods of \citet{lacy04} and \citet{stern05}, which
work very well to find luminous optically-unobscured sources, are not particularly efficient in finding obscured/low-luminosity AGN. Fig.~\ref{iracol}
shows two combinations of IRAC color-color diagrams for the X-ray sources and 24~$\mu$m-selected AGN candidates in the ECDF-S.
Interestingly, the X-ray detected sources with $f_{24}$/$f_R$$>$1000 and $R$-$K$$>$4.5 fall inside the AGN regions of \citet{lacy04} and \citet{stern05}.
However, the vast majority of the X-ray-undetected IR-red excess sources are found outside the \citet{stern05} region and on the edge
of the \citet{lacy04} locus.

As for the $f_{24}$/$f_8$ flux ratio, we also find in the IRAC colors significant differences between the X-ray detected and undetected samples.
In the IRAC case, most of the differences are found at longer wavelengths. In fact, we found very small differences in the [3.6]-[4.5] color 
(average [3.6]-[4.5]=0.79 for X-ray sources and 0.69 for the X-ray undetected sample), while significant differences can be seen in the
[5.8]-[8] color (average of 0.99 for X-ray sources versus 0.5 for the X-ray undetected sample). As at longer wavelengths, these differences 
can be well explained by the effects of self-absorption of the dust re-emission. Using the \citet{nenkova08a} torus models, we found that 
for heavily absorbed sources the emission at wavelength shorter than $\sim$5~$\mu$m is dominated by stellar light, as the AGN re-emission
is severely obscured. This explains why the 24~$\mu$m-selected sources not detected in X-rays fall in the same region of the [3.6]-[4.5] versus
[5.8]-[8] diagram as inactive galaxies at $z$$>$2 \citep{barmby06}. In the case of the X-ray-detected sources, the significant torus contribution 
in the observed-frame 8~$\mu$m band explains why these sources are displaced to the right in this diagram, so that they fall inside the \citet{stern05} 
diagram. A very similar effect can be seen in the right panel of Fig.~\ref{iracol}, where self-absorption can explain the observed difference
in $f_8$/$f_{4.5}$ flux ratio.

The mid-IR colors of 24~$\mu$m-excess sources were also studied by \citet{pope08} in the GOODS-N field, to similar MIPS 24-$\mu$m depths. The first 
surprising result is that they found that $>$90\% of their sources with $f_{24}/f_R$$>$1000 also have $R$-$K$$>$4.5. In contrast, we found that this 
fraction is 77\% in our sample, while \citet{fiore08} report a fraction of $\sim$50\%. One explanation mentioned by \citet{pope08} is that the 
fainter 24~$\mu$m sources have in general blue $R$-$K$ colors. However, in our sample we found that out of the 74 sources 
with $f_{24}$/$f_R$$>$1000 and $R$-$K$$<$4.5 only 8 have $f_{24\mu m}$$<$100~$\mu$Jy. Another possibility is that the different 
$K$-band magnitude limit of each sample can explain these discrepancies. While \citet{pope08} do not quote their flux limit in the $K$ band, they mention 
that a large fraction (75 of 79, 95\%) of the sources with $f_{24\mu m}$/$f_R$$>$1000 are detected in the $K$-band. Similarly, the $K$-band imaging in 
the GOODS-S region used by \citet{fiore08} is deep enough to detected all their 24~$\mu$m sources. Hence, it is unlikely that these differences are due 
to the $K$-band flux limit either.

The main conclusion of \citet{pope08} is that a large majority, 80\%, of their sources with $f_{24}$/$f_R$$>$1000 
(most of them also have $R$-$K$$>$4.5) are dominated by star-formation and not AGN activity. They base this conclusion in the low value
of $f_8$/$f_{4.5}$ (their Figure 4), which is taken as evidence for a mid-IR emission dominated by a star-forming galaxy. In our sample, a large
fraction of sources have $f_8$/$f_{4.5}$$<$2 (their threshold for star-formation versus AGN), including most of the X-ray detected 24~$\mu$m selected
sources, in which the AGN nature derived from the X-ray luminosity and spectral shape is clear. We point out that these low $f_8$/$f_{4.5}$ values can 
easily be explained by the effects of self-absorption in the 8~$\mu$m band, as can be seen using the \citet{nenkova08b} torus models, and as discussed 
above. \citet{pope08} also use the X-ray spectral shape to support their conclusion of a low number of AGN in their 24~$\mu$m-selected
sample. However, as we study in more detail in section \ref{x_stack} and was shown by \citet{fiore08}, proper simulations are required to convert
the observed X-ray spectral shape from a stacked sample that includes both AGN and star-forming galaxies into an AGN fraction.

\subsection{X-ray to Mid-IR Ratio and AGN Fraction}
\label{xray_midIR}

\begin{figure}
\begin{center}
\plotone{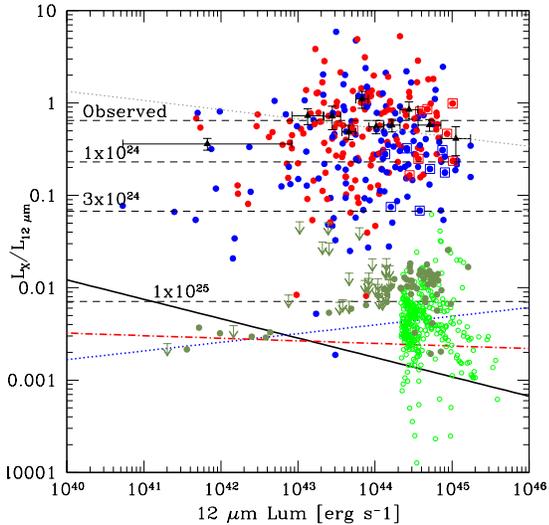}
\end{center}
\caption{Ratio of hard X-ray to mid-IR (at rest-frame 12~$\mu$m) versus rest-frame 12~$\mu$m luminosity. {\it Red circles} show the X-ray detected
sources with HR$>$-0.3, while the {\it blue circles} mark the sources with HR$<$-0.3. {\it Black triangles with error bars} show the average ratio for
the X-ray-detected sources in bins of 30 sources. Circles enclosed by squares identify the X-ray detected IR-red excess sources. The {\it dashed line} 
at a ratio of $\sim$0.6 shows the average value of $L_X$/$L_{12~\mu m}$ for all the X-ray sources. The {\it dotted line} shows the relation between
intrinsic X-ray and 12~$\mu$m luminosity for local Seyfert galaxies \citep{gandhi09}. {\it Green circles} show the location of the most 
luminous rest-frame 12~$\mu$m X-ray undetected sources outside the selection region. X-ray undetected IR-red excess sources 
are shown by olive circles and upper limits. X-ray luminosities were derived from the stacking signal.  {\it Dashed} lines marked 
``1$\times$10$^{24}$'', ``3$\times$10$^{24}$'' and ``1$\times$10$^{25}$'' show the effects of X-ray obscuration in the 
observed $L_X$/$L_{12 \mu m}$ ratio. The thick {\it solid, dot-dashed and dotted lines} show the expected $L_X/L_{12\mu m}$ for 
star-forming galaxies, considering different recipes to convert star formation rates into X-ray and IR luminosities (see text for details).}
\label{x_24_corr}
\end{figure}

Both X-ray and mid-IR are tracers of AGN activity, either direct emission in the former or re-radiation in the latter, and of recent star formation, although the 
relative X-ray emission is significantly lower in the latter. Hence, the X-ray to mid-IR ratio is particularly important in separating emission from AGN 
and star formation (e.g., \citealp{horst06,ramos-almeida07}). In  Fig.~\ref{x_24_corr} we show the hard (observed-frame 2--8~keV) X-ray versus 
mid-IR (at rest-frame 12~$\mu$m) luminosity ratio as a function of rest-frame 12~$\mu$m luminosity for the X-ray sources in the ECDF-S. The 
rest-frame 12~$\mu$m fluxes were obtained from the quasar template spectrum of \citet{richards06}, normalizing it to the observed MIPS 24~$\mu$m flux, 
which at the typical redshifts of our sources is very similar to rest-frame 12~$\mu$m. This particular wavelength was chosen since it is nearly 
unaffected by obscuration (e.g., \citealp{treister09a}) and it was also used by previous studies of the local AGN population (e.g., \citealp{horst06,gandhi09}), thus
allowing for a direct comparison with our high-redshift sample. Typical values of the conversion from observed-frame 24~$\mu$m to 
rest-frame ~12$\mu$m are $\sim$0.91-1.26, with an average of 1.0. 

The average X-ray to mid-IR ratio for the X-ray sources is 0.65, while  the median is 0.4. As can be seen in Fig.~\ref{x_24_corr} this ratio for the 
X-ray detected AGN is spread over one order of magnitude. One advantage of this ratio is that, at least for Compton-thin absorption levels, it is 
independent of obscuration. In fact, insignificant differences were found for the average X-ray to mid-IR ratio values for obscured and unobscured 
sources in the X-ray detected sample. However, for more obscured sources, this ratio is expected to decrease, due to the effects of absorption in 
the hard X-ray bands. For example, for the 24~$\mu$m-selected sources individually detected in X-rays, which typically have 
$N_H$$\simeq$10$^{23}$~cm$^{-2}$, the average X-ray to mid-IR ratio is 0.43.

In order to model the effects of obscuration in the hard X-ray to mid-IR ratio for CT sources, both photoelectric absorption and Compton scattering
need to be taken into account. We have done this using the template X-ray spectra of \citet{matt99a}, computed using Monte Carlo simulations.
For a heavily obscured source with $N_H$$>$3$\times$10$^{24}$~cm$^{-2}$ the observed ratio decreases by 1--2 orders of magnitude.
In the case of the IR-red excess sources which are not individually detected in X-rays we computed X-ray fluxes from the measured stacked
signal. For each source, the average X-ray flux measured in 24~$\mu$m flux bins as described in section \ref{x_stack} was assigned. For the fainter
24~$\mu$m sources, for which no X-ray stacked signal was measured, the 1-$\sigma$ background fluctuations are used as upper limits. Very similar 
and consistent results are obtained if the X-ray stacking is done binning sources based on 24~$\mu$m luminosity instead.

As can be seen in Fig.~\ref{x_24_corr}, the X-ray to mid-IR ratio for the IR-red excess  sources is about two orders of magnitude smaller
than the average value for the X-ray-detected sample and for most sources falls in the range expected for obscuration of $\sim$5$\times$10$^{24}$ to
10$^{25}$~cm$^{-2}$. A small fraction of the IR-red excess sample, 20\%, have X-ray to mid-IR ratios lower than 7$\times$10$^{-3}$. It would be very hard to explain 
such low ratios using obscuration, and hence it is more likely that these sources are not AGN but star-forming galaxies. On the other hand, the observed
ratio for the IR-red excess sample is significantly larger than for the sources outside our selection region, i.e. $f_{24}$/$f_R$$<$1000 or $R$-$K$$<$4.5,
even at similar rest-frame 12~$\mu$m luminosities. The observed X-ray to mid-IR ratio for IR-red excess sources at 
$L_{12\mu m}$$\sim$10$^{45}$~erg~s$^{-1}$ is roughly 10 times larger than the values expected from the X-ray versus mid-IR luminosity for star-forming 
galaxies reported by \citet{donley08}. Similar results are found if the relation between star formation rate and X-ray luminosity proposed 
by \citet{gilfanov04}, which is consistent with the \citet{ranalli03} correlation, is used together with the relationship between IR luminosity and star formation 
rate of \citet{kennicutt98}. In order to transform rest-frame 12~$\mu$m into total IR luminosity we used the observed correlations given by \citet{chary01} and 
by \citet{takeuchi05}. The obtained rest-frame 2--10~keV luminosities are converted into observed-frame 2--8~keV assuming $<$$z$$>$=2 and the typical 
X-ray spectrum of high-mass X-ray binaries (e.g., \citealp{lutovinov05}), namely $\Gamma$=1.0 and a high energy cutoff, $E_c$, of 20~keV. As can be 
seen in Fig.~\ref{x_24_corr}, {\it the X-ray to mid-IR ratio for the IR-red excess sources is significantly higher than the expected value for pure 
star-forming galaxies, thus confirming the AGN nature for the vast majority of our sample}. 

\begin{figure}
\begin{center}
\plotone{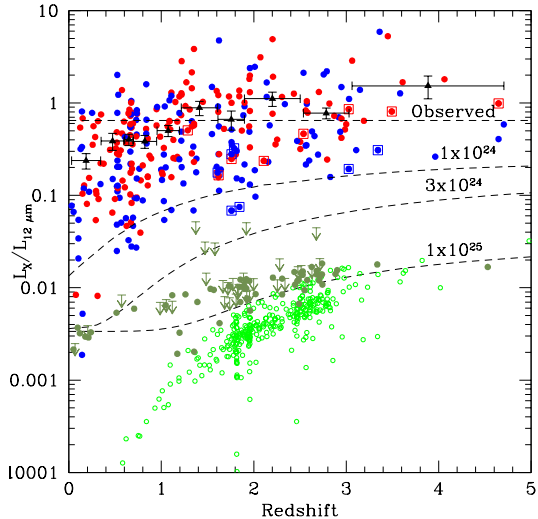}
\end{center}
\caption{Ratio of hard X-ray to mid-IR ratio as a function of redshift. Symbols are the same as in Fig.~\ref{x_24_corr}. The dependence 
of this ratio on redshift for a constant obscuration value is shown by the {\it dashed lines}. As can be seen here, the IR-red excess sources
({\it olive circles}) are mostly located in the region expected for AGN with obscurations $>$5$\times$10$^{24}$~cm$^{-2}$ at all redshifts
and are systematically above the sample of sources with similar IR luminosities but bluer $f_{24\mu m}$/f$_{R}$ colors. Based on this figure
we can conclude that the redshift spread of our sample does not affect our conclusions. }
\label{x_24_red}
\end{figure}

To investigate a possible dependence on redshift, in Fig.~\ref{x_24_red} we show the X-ray to mid-IR ratio as a function of redshift.
As can be seen in this figure, the IR-red excess sources are systematically above the sample with similar IR luminosities and bluer 
$f_{24~\mu m}$/$f_R$ colors. At the same time, the IR-red excess sources are located in the region expected for heavily obscured AGN
with $N_H$$>$3$\times$10$^{24}$~cm$^{-2}$ at all redshifts. This confirm that our conclusions are not affected by the redshift spread in
our sample. It is also interesting that the fraction of star-forming galaxies in the IR-red excess sample inferred from the X-ray to mid-IR ratio 
is consistent with the value found from the comparison of the observed HR with the simulations of \citet{fiore09}.  While some trend is apparent, the 
evidence for a possible dependence of the fraction of AGN-dominated sources on mid-IR luminosity is not conclusive in our sample. 
Quantitatively, the fraction of star-forming galaxies (measured using a X-ray to mid-IR ratio of 7$\times$10$^{-3}$ as the dividing point 
between AGN and star-forming galaxies) in the 10$^{43}$$<$ $L_{12\mu m}$ (erg~s$^{-1}$)$<$10$^{44}$ luminosity range is 
4/17=23.5$\pm$20\%, considering only statistical uncertainties, while for sources with $L_{12\mu m}$$>$10$^{44}$~erg~s$^{-1}$ is 
6/64=9.4$\pm$6\%. Furthermore, for sources with $L_{12\mu m}$$<$10$^{43}$erg~s$^{-1}$ is 8/9=89$^{+11}_{-60}$\%. The number of sources is 
small for lower luminosity sources ($L_{12\mu m}$$<$10$^{43}$erg~s$^{-1}$), mainly due to a combination of the MIPS flux limit and our 
selection method. Using the same threshold in X-ray to mid-IR ratio, the fraction of star-forming galaxies for sources 
with $L_{12\mu m}$$>$10$^{44}$erg~s$^{-1}$ and outside our obscured AGN selection region is 80$\pm$7\%, thus showing the clear 
differences in X-ray properties for our 24~$\mu$m selected sample, even at the same mid-IR luminosities.

\subsection{Space density of CT AGN}

Given the strong evidence presented here for the AGN nature of the majority of the IR-red excess sources, and the very high obscuration levels
that these sources show, we can use this sample to study the properties of the CT AGN population at $z$$\gtrsim$2. While CT AGN can be abundant 
at high redshift, $z$$>$2, they only represent a small contribution to the extragalactic X-ray background, roughly 1-2\%, as was shown 
by \citet{treister09b}. Because of the current uncertainties in the measurements of this background radiation and degeneracies in the assumed models, this 
integral constraint cannot be used to infer the number of heavily obscured sources at high redshift. While \citet{treister09b} discuss a possible stronger 
evolution for heavily-obscured AGN relative to the less-obscured sources, \citet{dellaceca08a} concluded that the luminosity functions of X-ray-selected 
CT and less-obscured AGN were consistent with each other. These potentially contradictory results are studied in more detail here.

We measured the space density of CT AGN using our sample of IR-red excess sources. Because of the potential incompleteness of
our photometric redshift determinations we restrict our study to $z$$\leq$3. Furthermore, taking into account the flux limit of the MIPS 24~$\mu$m 
data in the ECDF-S, the space density was measured only at $z$$<$1.1 for sources with $L_{24\mu m}$$<$10$^{44}$erg~s$^{-1}$. These
choices minimize the effects of incompleteness in our sample.  However, other sources of incompleteness, like the $K$-band magnitude limit,
required for our color selection could be also important. X-ray luminosities were estimated from the stacked X-ray signal, as described in 
Section \ref{xray_midIR}. 

\begin{figure}
\begin{center}
\plotone{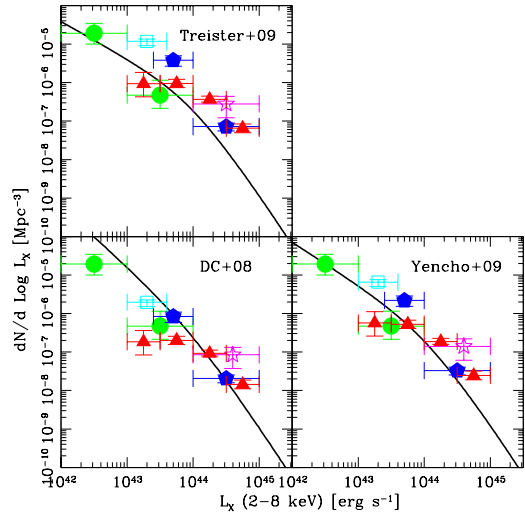}
\end{center}
\caption{Three estimates ({\it thick lines}) of the hard X-ray luminosity function of Compton thick AGN at $z$=0, as measured by 
\citet{treister09b}  ({\it upper left panel}),  Della Ceca et al. (2008a; {\it lower left panel}), and Yencho et al. (2009; {\it right panel}). Symbols with
error bars show the measurements of the space density of Compton thick AGN as a function of hard X-ray luminosity, de-evolved to $z$=0
following the corresponding evolution for each work (see text for details). {\it Filled triangles:} this paper, {\it filled circles}: \citet{treister09b}, 
{\it filled pentagons:} \citet{fiore09}, {\it open square}: \citet{daddi07}, {\it star}: \citet{alexander08}. }
\label{dens_lum}
\end{figure}

In Fig.~\ref{dens_lum} we present our measurements of the space density as a function of luminosity (triangles) together 
with the expectations from three different $z$=0 CT AGN luminosity functions: \citet{dellaceca08a}, \citet{treister09b} and \citet{yencho09}. For
the latter, the X-ray luminosity function for all AGN was converted into a CT AGN luminosity function assuming the luminosity dependence of the
obscured AGN fraction given by \citet{treister09b}, normalized to match the INTEGRAL measurements at low luminosities. Except for the lack of a 
decline at low luminosities in the work of \citet{dellaceca08a}, the three curves are consistent with each other. In order to compare
with these $z$=0 expectations, the ECDF-S measurements, together with other values obtained from the literature were ``de-evolved'' using the 
corresponding evolution form for each luminosity function. The differences between observations and expectations must be due to the assumed 
evolution in the latter. While in all these works a luminosity-dependent density evolution was assumed, different parameter values were obtained in each 
case. For example, \citet{dellaceca08a} assumed a strong evolution for all AGN, while \citet{treister09b} used the values estimated by \citet{ueda03}, which 
give a much shallower evolution. As can be seen in this figure, while both the \citet{dellaceca08a} and \citet{yencho09} provide an acceptable description 
of the observational data, the values obtained using the \citet{ueda03} evolution fall systematically above the $z$=0 curve. Hence, these results indicate that 
a similar shape of the luminosity function is found for CT AGN and less-obscured sources, in agreement with the conclusions of \citet{dellaceca08a}.

\begin{figure}
\begin{center}
\plotone{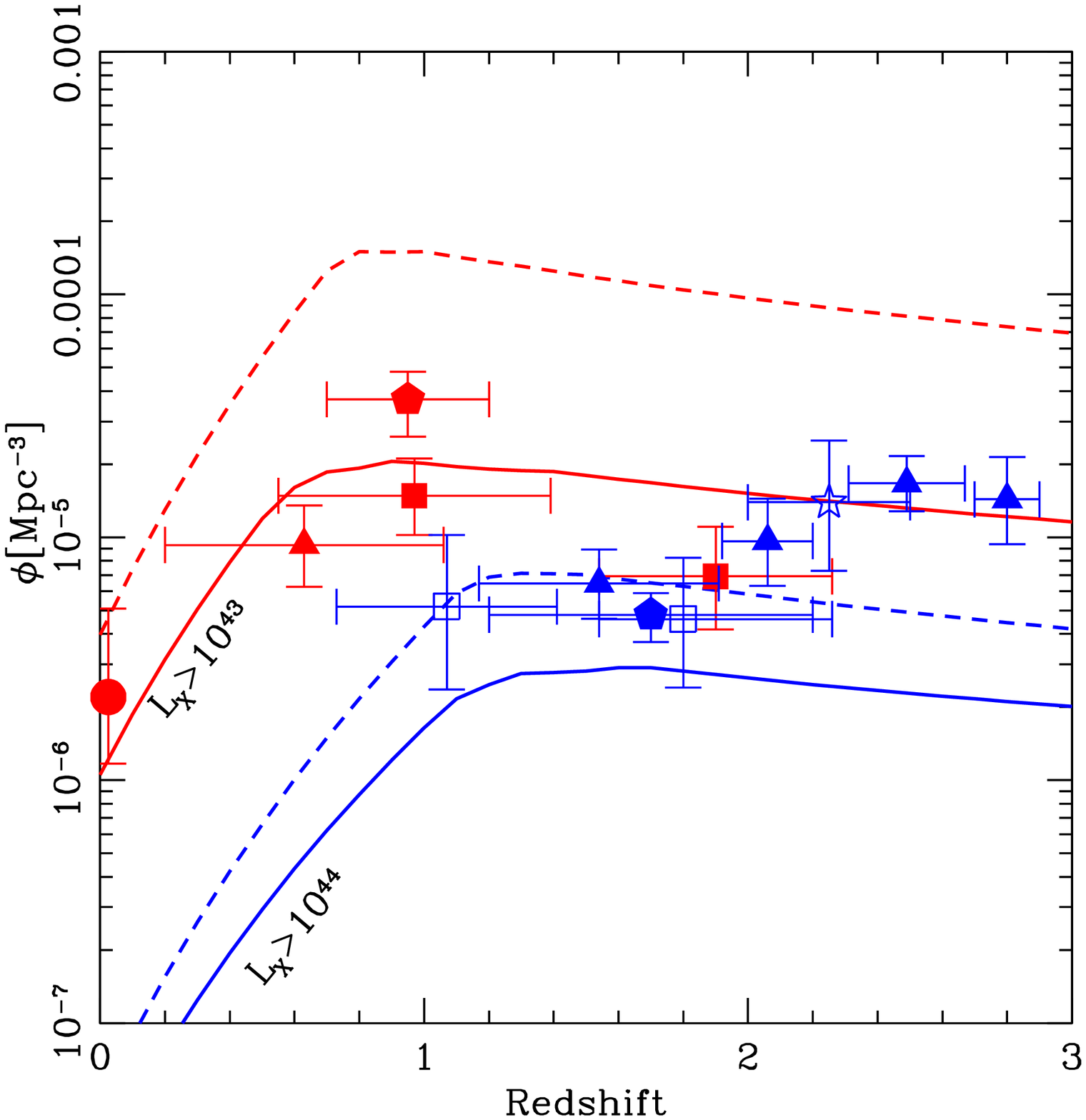}
\end{center}
\caption{Space density of Compton thick AGN as a function of redshift. {\it Filled triangles} show the measurements described in this paper. {\it Squares}: 
space density from the work of \citet{tozzi06}. {\it Star}: Measurement by \citet{alexander08}. {\it Pentagons}: Values reported by \citet{fiore09}. {\it Solid
lines} show the expected space density of Compton thick AGN from the luminosity function of \citet{yencho09}, with the overall normalization fixed to the 
results of \citet{treister09b}, while the {\it dashed lines} show the expectations based on the luminosity function of \citet{dellaceca08a}. {\it Red symbols} show 
measurements and expectations for $L_X$$>$10$^{43}$~erg~s$^{-1}$ sources, while the {\it blue symbols} 
are for $L_X$$>$10$^{44}$~erg~s$^{-1}$. While for the lower luminosity sources a good agreement is found between observations and expectations, higher 
luminosity sources lie well above the luminosity function. Furthermore, the new measurements seem to indicate as strong increase in the number of 
high-luminosity heavily-obscured sources at $z$$>$2.}
\label{com_dens}
\end{figure}

In order to study in more detail the evolution of the CT AGN space density, in Fig.~\ref{com_dens} we present our measurements of
the CT AGN space density as a function of redshift. A reasonable agreement, in particular at $z$$<$1, is found between both
our observed values and others obtained from the literature and the luminosity function and evolution assumed by \citet{yencho09}.
However, at higher redshifts and higher luminosities clear discrepancies are found. This was already pointed out by \citet{treister09b}, who
concluded that this difference of a factor of 2--3 could most likely be due to either incompleteness in the Swift/BAT and INTEGRAL CT AGN
samples at $z$=0 used to fix the luminosity function normalization (because reflection-dominated AGN are missed) or to contamination 
by other types of sources in the observed values at high redshifts. However, after adding the measurements obtained using the IR-red excess 
sources in the ECDF-S it appears not only that the systematic difference is still present but perhaps more importantly that there is a strong 
increase in the number of CT AGN from $z$$\simeq$1.7 to 2.4.

As can be seen in Fig.~\ref{x_24_corr}, most of the sources with only upper limits to their X-ray to mid-IR emission ratios have mid-IR 
luminosities lower than 10$^{44}$~erg~s$^{-1}$. Hence, excluding these sources from our measurements of the space density of CT AGN
as a function of redshift presented in Fig.~\ref{com_dens} does not change these values significantly. For sources with 
$L_X$$>$10$^{43}$~erg~s$^{-1}$ if those with only upper limits are excluded the space density is reduced by $\sim$30\%. In Fig.~\ref{dens_lum},
where we presented the space density of CT sources as a function of hard X-ray luminosity, the effects of excluding sources with upper
limits are more relevant for the less-luminous sources. In the 43$<$log($L_X$ [erg~s$^{-1}$])$<$43.5, the space density is reduced by
$\sim$70\% if upper limit sources are excluded. Similarly,  in the 43.5$<$log($L_X$ [erg~s$^{-1}$])$<$44, the space density should be 
reduced by $\sim$60\% by excluding the upper limits from the sample, while at higher luminosities the effect is less than 10\% and hence
negligible. 

The strong evolution in the number of CT AGN at high redshift presented in Fig.~\ref{com_dens} was not indicated by any existing luminosity 
function. It is unlikely that this evolution is due to selection effects, as results from different fully independent surveys and selection techniques 
are combined in Fig.~\ref{com_dens}, namely X-ray selected sources from \citet{tozzi06}, 24~$\mu$m-selected sources from \citet{fiore09} and 
our work, and the sample of CT AGN found using mid-IR spectroscopy by \citet{alexander08}.  If confirmed, this result indicates a very large number
of heavily obscured AGN at $z$$\simeq$2.5 followed by a strong decline, by a factor of $\sim$3, at $z$=1.5. This result can be interpreted
in the context of the galaxy evolution models of \citet{hopkins08}, where quasar activity is driven by galaxy mergers and the supermassive
black hole is initially completely surrounded by dust, before radiation pressure removes it and a ``classical'' unobscured quasar is visible.
The consequences of this observed rapid evolution are beyond the scope of this work and will be studied in further detail in an upcoming paper.

\section{Conclusions}

We presented a study of heavily-obscured (CT) AGN candidates  selected from their optical, near and mid-IR colors. These sources
were selected by requiring $f_{24\mu m}$/$f_R$$>$1000 and $R$-$K$$>$4.5. Using this selection method, we found a total of 211 
infrared-red excess sources in the ECDF-S. Of these, 18 were individually-detected in the Chandra 250-ksec observations of this field.  These
sources are moderately-obscured, with $N_H$=10$^{22}$-10$^{23}$~cm$^{-2}$, according to their X-ray spectra. Two of
these sources have $N_H$$>$10$^{-24}$~cm$^{-2}$, and therefore are classified as CT AGN, including one at $z$=4.65.
The average redshift for the X-ray-detected sample is $<$$z$$>$$\sim$2.37 (median=1.85), in agreement with previous studies. By performing 
spectral fitting to the observed optical to near-IR fluxes of the X-ray-detected sources we found that the host galaxies have typical 
stellar masses of 10$^{11.7}$M$_\odot$ and are subject to moderate extinction, with $<$E(B-V)$>$=0.6.

For the X-ray undetected sources, we found a slightly lower average redshift of $<$$z$$>$$\sim$1.9, based on photometric measurements
only. We found that in general the X-ray-detected sources are brighter in both the 24~$\mu$m and optical bands compared to
the undetected ones. Taking advantage of the low Chandra background, we performed X-ray stacking, finding a significant
($\gtrsim$3$\sigma$) detection in both the soft and hard bands. The average X-ray flux of these sources is 
$\sim$8$\times$10$^{-17}$~erg~cm$^{-2}$s$^{-1}$, indicating that in order to detected these sources individually, exposure
times of $\sim$10 Msec with Chandra would be required. Taken at face value, the observed average HR of 0.13 corresponds to
a $N_H$ of $\sim$2$\times$10$^{23}$~cm$^{-2}$. However, this can also be interpreted as the effects of combining a population of 
90\% CT AGN and 10\% star-forming galaxies, which in general have softer X-ray spectra. From this analysis we found marginal
evidence for a small dependence of the fraction of AGN relative to star-forming galaxies on 24~$\mu$m flux, going from $\sim$95\%
for the brightest sources to $\sim$80\% at the lowest flux bin. The spectral fitting analysis performed to these sources indicate 
that in general there is evidence for substantial young stellar populations, younger than 100 Myrs. This suggests that these sources are
simultaneously experimenting significant star-formation and heavily-obscured AGN activity. We did not found significant differences in the 
stellar masses and extinction values for the host galaxies of the X-ray detected and undetected IR-red excess sources.

The X-ray undetected IR-red excess sources have redder $f_{24}$/$f_8$ flux ratios than those detected in X-rays.
The average $f_{24}$/$f_8$$\simeq$25 is hard to explain assuming the observed spectra of ultra-luminous infrared galaxies,
however it can be well-explained by dust re-radiation models in which the effects of self-absorption are important. Similarly, while 
the IRAC colors of most of these sources are outside the typical AGN region, this can be explained by the effects of absorption and/or
a dominance of the near-IR emission by the host galaxy. In contrast, the X-ray detected sources have IRAC colors consistent with
those of less-obscured AGN. The X-ray to mid-IR ratio for the X-ray-detected sources is similar to the average value for the overall
X-ray population. For the X-ray-undetected sources this ratio is $\sim$2 orders of magnitude lower than for the sample individually
detected in X-rays. This ratio is consistent with obscuration levels of $\sim$5$\times$10$^{24}$ to 10$^{25}$~cm$^{-2}$, while it
is significantly larger than the relative X-ray emission expected from star-formation activity alone. Using a constant threshold
in X-ray to mid-IR flux ratio, we found that the fraction of AGN in our sample should be  $\sim$80\%, consistent with the value found
from the analysis of the stacked X-ray signal. We also found a similar dependence of the AGN fraction on 24~$\mu$m luminosity
ranging from $\sim$90\% at the bright end to only $\sim$10\% for the faintest sources.

Finally, we studied the space density of sources implied by our sample of 24~$\mu$m-selected heavily-obscured AGN candidates. While at 
lower redshifts and luminosities we found a good agreement with population synthesis models and extrapolations of existing X-ray 
luminosity functions to higher obscuration, we found significant excesses for high-luminosity sources at high redshifts. Furthermore, we 
found a strong evolution in the number of sources with $L_X$$>$10$^{44}$~erg~s$^{-1}$ from $z$=1.5 to 2.5. This can be interpreted 
as evidence for a relatively short-lived heavily-obscured phase before the strong radiation pressure removes the surrounding dust and 
turns the sources into an unobscured quasar.

\acknowledgements

We thank Moshe Elitzur, Maia Nenokva and Robert Nikutta for providing us their CLUMPY models in electronic format, and Takamitsu Miyaji for
his support of the CSTACK tool. We are grateful
for the help of Youichi Ohyama reducing the MOIRCS data and for useful discussions with Dave Sanders, Priya Natarajan, David Rupke,
Len Cowie and Amy Barger. We thank the anonymous referee for very helpful comments. Support for the work of ET was provided by the
National Aeronautics and Space Administration through Chandra
Postdoctoral Fellowship Award Number PF8-90055 issued by the Chandra
X-ray Observatory Center, which is operated by the Smithsonian
Astrophysical Observatory for and on behalf of the National
Aeronautics Space Administration under contract NAS8-03060. EG acknowledges support from NSF grant AST-0807570. This 
research has made use of NASA's Astrophysics Data System.


\end{document}